\documentclass[twocolumn,smallextended]{svjour3}      
\pdfoutput=1
\usepackage{svg}
\usepackage{enumitem}
\usepackage{scrextend}
\usepackage{amsmath,amssymb,amsfonts}
\usepackage{algorithmic}
\usepackage{graphicx}
\usepackage{textcomp}
\usepackage{xcolor}
\usepackage{longtable}
\usepackage{float}
\usepackage{graphicx}
\usepackage{multirow}
\usepackage{balance}
\usepackage{hyperref}
\usepackage{tabularx} 
\usepackage{booktabs} 
\usepackage{hhline}
\usepackage{array}
\usepackage{cite}
\newcolumntype{L}[1]{>{\raggedright\let\newline\\\arraybackslash\hspace{0pt}}m{#1}}
\newcolumntype{C}[1]{>{\centering\let\newline\\\arraybackslash\hspace{0pt}}m{#1}}
\newcolumntype{R}[1]{>{\raggedleft\let\newline\\\arraybackslash\hspace{0pt}}m{#1}}

\usepackage[most]{tcolorbox}
\usepackage{changepage}
\usepackage{dblfloatfix}
\usepackage{multirow}
\usepackage{colortbl}
\usepackage{lscape}

\usepackage{caption}
\usepackage{subcaption}

 \definecolor{darkgreen}{rgb}{0.0, 0.2, 0.13}
 
 \definecolor{babyblue}{rgb}{0.54, 0.81, 0.94}
\usepackage{pgfplotstable}
    \usetikzlibrary{
        pgfplots.colorbrewer,
    }
    
    \pgfplotsset{ticks=none,
colormap/RdBu-5,
            cycle list={
                {index of colormap={2},fill=.,draw=.},
                {index of colormap={1},fill=.,draw=.},
                {index of colormap={0},fill=.,draw=.},
                {index of colormap={2},fill=.,draw=.},
                {index of colormap={3},fill=.,draw=.},
                {index of colormap={4},fill=.,draw=.},}}

\newcommand{\surveyquote}[1]{\begin{addmargin}[1em]
{0em}\emph{#1}\end{addmargin}}
\newcommand{\tickerRating} [5]{ 
 \resizebox {2.4cm} {0.75cm} {
\begin{tikzpicture}
\begin{axis}[
      title=\empty,
      xlabel=\empty,
      ylabel=\empty, axis lines=none,
      yticklabels=\empty,
     bar width=0.8cm , ymin=0, ymax=72,   
    ]
    \addplot[ybar, fill=babyblue] coordinates {
        (1,{#1}) (2,{#2}) (3,{#3})( 4,{#4})(5,{#5})
    };
    \end{axis}

\end{tikzpicture}
}}

\newcommand{\tickerRatingTeal} [5]{ 
 \resizebox {2.4cm} {0.75cm} {
\begin{tikzpicture}
\begin{axis}[
      title=\empty,
      xlabel=\empty,
      ylabel=\empty, axis lines=none,
      yticklabels=\empty,
     bar width=0.8cm , ymin=0, ymax=72,   
    ]
    \addplot[ybar, fill=darkgreen] coordinates {
        (1,{#1}) (2,{#2}) (3,{#3})( 4,{#4})(5,{#5})
    };
    \end{axis}

\end{tikzpicture}
}}

\journalname{Empirical Software Engineering}
\definecolor{blue_}{HTML}{B8EBFF}
\definecolor{pink_}{HTML}{FFCCCB}

\newcommand{\code}[1]{\texttt{ #1}}

\newcommand{\revision}[1]{\textcolor{black}{#1}}

\newcommand{\positive}[1]{\cellcolor{blue_}{#1}}
\newcommand{\negative}[1]{\cellcolor{pink_}{#1}}

\newcommand{\bluebg}[1]{\colorbox{blue_}{#1}}
\newcommand{\pinkbg}[1]{\colorbox{pink_}{#1}}

\newenvironment{boxedtext}
    {
    
    \begin{center}

    \begin{tabular}{|p{0.96\linewidth}|}
    \hline
    }
    { 
    \\ \hline
    \end{tabular} 
    
    \end{center}
       }

\begin{document}

\title{What Makes a Code Review Useful to OpenDev Developers? An Empirical Investigation}

\author{   Asif Kamal Turzo  \and
        Amiangshu Bosu
}


\institute{A. Turzo, and A. Bosu \at
              Department of Computer Science, Wayne State University, Detroit, Michigan. \\
              \email {asifkamal@wayne.edu}, {amiangshu.bosu@wayne.edu}           
          }

\date{Submitted: February 20, 2023 / Accepted: TBD}
\titlerunning{What Makes a Code Review Useful to OpenDev Developers?}
\authorrunning{Turzo and Bosu}


\maketitle

\begin{abstract}

\textit{Context: }
Due to the association of significant efforts, even a minor improvement in the effectiveness of Code Reviews(CR) can incur significant savings for a software development organization. 

\textit{Aim:} This study aims \textit{ to develop a finer grain understanding of what makes a code review comment useful to  OSS developers, to what extent a code review comment is considered useful to them,  and how various contextual and participant-related factors influence its degree of usefulness}. 

\textit{Method:}
On this goal, we have conducted a three-stage mixed-method study. We randomly selected 2,500 CR comments from the OpenDev Nova project and manually categorized the comments. We designed a survey of OpenDev developers to understand their perspectives on useful CRs better. Combining our survey-obtained scores with our manually labeled dataset, we trained two regression models - one to identify factors that influence the usefulness of CR comments and the other to identify factors that improve the odds of `Functional' defect identification over the others. 

\textit{Key findings:}
 The results of our study suggest that a CR comment's usefulness is dictated not only by its technical contributions, such as defect findings or quality improvement tips, but also by its linguistic characteristics, such as comprehensibility and politeness. While a reviewer's coding experience positively associates with CR usefulness, the number of mutual reviews, comment volume in a file, the total number of lines added /modified, and CR interval have the opposite associations. 
While authorship and reviewership experiences for the files under review have been the most popular attributes for reviewer recommendation systems, we do not find any significant association of those attributes with CR usefulness. 

\textit{Conclusion:} We recommend discouraging frequent code review associations between two individuals as such associations may decrease CR usefulness. We also recommend authoring CR comments in a constructive and empathetic tone.  As several of our results deviate from prior studies, we also recommend more investigations to identify context-specific attributes to build reviewer recommendation models.


\end{abstract}

\section{Introduction}
\label{sec:intro}
Code Review (CR) is a software development practice where developers asynchronously inspect peers' code changes to identify defects as well as potential quality improvement opportunities~\cite{bacchelli2013expectations}.
 By addressing the cost ineffectiveness of traditional Fagan inspections~\cite{fangan1976}, CR  has achieved widespread adoption among commercial and Open Source Software (OSS) development organizations~\cite{bacchelli2013expectations,rigby2013convergent,sadowski2018modern}.
 Many projects mandate each code change be approved through a CR process before it can be integrated into the project's main codebase~\cite{rigby2011understanding}. As a result, both commercial and OSS developers are spending 10-15\% of their time on CR tasks~\cite{bosu2016process}.
 Due to the large efforts associated with CRs, even a minor improvement in CR's effectiveness can save developers time and can incur significant savings for a software development organization. 
Since prior studies report between 20 to 44\% of CRs are marked as `not useful' by code authors~\cite{bosu2015characteristics,rahman2017predicting,hasan2021using}, improving CR effectiveness is a high priority for many organizations~\cite{bosu2015characteristics,hasan2021using}. 

The basic building blocks of a CR are suggestions (aka review comments) authored by the reviewers. A CR is effective if comments belonging to that CR are useful to the code author and involved reviewers~\cite{bosu2015characteristics}. Therefore, an empirical understanding of what makes CR comments useful to code authors and reviewers is essential to improve CR effectiveness. \revision{Among the various approaches to improve CR effectiveness, which include adding features to CR tools~\cite{henley2018cfar,barnett2015helping,tao2015partitioning}, automated reviews~\cite{balachandran2013reducing,tufan2021towards,tufano2022using}, reviewer recommendation~\cite{zanjani2015automatically,thongtanunam2015should}, and promoting useful feedback~\cite{bosu2015characteristics,kononenko2015investigating}, this study focuses on the latter. }
Findings of recent studies investigating CR usefulness suggest that although a developer's primary expectations from a CR comment are the identification of defects, code improvement opportunities, or an alternative solution approach~\cite{bacchelli2013expectations,bosu2016process}, the majority of the CR  do not meet those expectations~\cite{beller2014modern,bosu2015characteristics,bacchelli2013expectations,czerwonka2015code}. A CR comment may still be considered useful, even though it did not meet any of the primary expectations, but satisfies other secondary criteria, such as improving code maintainability, facilitating knowledge sharing, or helping relationship formed between the participants~\cite{bacchelli2013expectations,bosu2016process}. Although developers have limited disagreements regarding the primary criteria to judge a CR's usefulness, the same cannot be said for those secondary criteria.  For example, some developers consider suggestions to improve code documentation or code visualization as useful, while others consider the opposite~\cite{hasan2021using}.  Therefore, questions remain about how those secondary criteria compare to each other among a broader pool of developers.

We have identified three knowledge improvement areas based on existing works investigating CR usefulness~\cite{bosu2015characteristics,han2021understanding,bacchelli2013expectations,kononenko2018studying,sadowski2018modern}.
\revision{First, prior studies on CR usefulness \revision{mostly} focused on commercial organizations~\cite{bacchelli2013expectations,bosu2015characteristics,hasan2021using,rahman2017predicting,sadowski2018modern,kononenko2018studying}, except Kononenko \textit{et} al., which investigated CR usefulness among Mozilla developers~\cite{kononenko2015investigating}. Therefore, we have limited insights regarding OSS developers' perspectives on useful CRs.}
Second, we have a limited understanding of how various contextual and participant factors may influence  CR usefulness. Although Bosu \textit{et} al. analyzed how several changeset and reviewer characteristics influence CR usefulness~\cite{bosu2015characteristics}, several crucial factors (i.e., detailed in Table~\ref{table:usefulness-factors}) were missing from their investigation.    These missing insights can be helpful in training reviewers, preparing code changes for CRs, and selecting the best reviewer for those changes.
Finally, to analyze CR usefulness factors, Bosu \textit{et} al. ~\cite{bosu2015characteristics} grouped code reviews into two categories, either `Useful' or `Not useful'. However, such a coarse grain analysis fails to identify crucial insights since a CR suggesting a `variable renaming' falls into the same bin as the one finding a `critical defect'. Recommendations obtained through this coarse-grained analysis are prone to amplifying trivial issues such as naming conventions and documentation since the majority of CR comments belong to that categories~\cite {bosu2015characteristics,hasan2021using,beller2014modern,Panichella2020empirical}.
As current recommendations on conducting CRs, as well as code reviewer recommendation systems, are designed to repeat past histories rather than to maximize bug finding~\cite{gauthier2021historical,zanjani2015automatically,rahman-correct}, a large number of functional defects escape CRs~\cite{czerwonka2015code,paul2021security}.

To fill in these three knowledge gaps, the primary objective of this study is  \textit{ to develop a finer grain understanding of what makes a code review comment useful to  OSS developers, to what extent a code review comment is considered useful to them,  and how various contextual and participant-related factors influence its degree of usefulness}. 
 On this objective, we designed a mixed-method three-step case study of the OpenDev community (formerly known as OpenStack). 
In the first step, we randomly selected 2,500 code review comments from the OpenDev Nova project and manually categorized those comments to identify the frequency of different categories of review comments. In the second step, we designed an online survey using samples from the first step to better understand developers' views regarding the usefulness of various categories of CR comments.
We distributed the survey among the OpenDev developers and obtained 160 usable responses. In the third step, we combined the insights obtained from the survey with the manually categorized dataset to develop two regression models to identify underlying factors that are associated with CR usefulness as well as the types of CR comments being authored. 
This study's design is motivated by Bosu \textit{et} al., as this study has overlapping objectives with theirs~\cite{bosu2015characteristics}. However,  this study's final two steps, as well as our analysis approach, significantly differ from them.

 Primary contributions of this study include:
\begin{itemize}
  \item A finer-grained understanding from OSS developers' points of view regarding what makes a code review useful;
  \item An empirically developed ranking of code review comment categories based on perceived usefulness;
  \item An empirical evaluation of how various contextual and participant-related factors are associated with CR usefulness.
  \item An empirical evaluation of the factors influencing receiving  `functional defect finding' comments vs not receiving such ones.
  \item Recommendations for practitioners to improve the code review process.
  \item To promote replication, our analysis scripts and aggregated dataset are publicly available at \\ \url{https://github.com/WSU-SEAL/CR-usefulness-EMSE}. 
\end{itemize}

The remainder of the paper is organized as the following.  
Section~\ref{sec:background} provides a brief overview of the research context.
Section~\ref{sec:reseach-questions} introduces our two research questions based on our primary objective.
Section~\ref{sec:methods} details our research methodology.
Section~\ref{sec:results-rq1} and ~\ref{sec:results-rq2} present the results of our first and second research questions, respectively.
Section~\ref{sec:implications} and ~\ref{sec-related-works}  discuss the key implications of our findings and related works, respectively. 
Section~\ref{sec:threats} and Section~\ref{sec:conclusion} discuss threats to validity and concludes the paper, respectively.

\section{Background}
\label{sec:background}
This section presents a brief overview of the contemporary CR process and regression analysis concepts that would be essential to understand our research method.

 \subsection{Code Review}

 \revision{Code review is the process of reviewing and rewriting a piece of code iteratively before merging it into the main codebase.} Contemporary code reviews are lightweight, asynchronous, tool-based, and time efficient. 
 Popular CR tools include  Gerrit, Phabricator, ReviewBoard, GitHub pull request, Critique, and CodeFlow. While these tools vary based on the set of supported features, the primary workflow is tool agnostic. 

 To start a CR process, a code author uploads code to a repository and creates a review request with a description detailing the CR's goal. The author can invite reviewer(s). Additional contributors with required access can self-assign to review as well. 
 CR tools support reviewers' comprehension through various features, such as a side-by-side view highlighting the differences between the current version and the previous one, the commit history of the file, and potential conflicts.
 Reviewers can raise concerns regarding a particular code segment by adding inline comments (e.g., Figure~\ref{fig:examples}). 
 The developer (author of the patch) can view the review comments and interact with the reviewers using the same interface. The author can address the comment by modifying the files and uploading a new patch set, or adding responses. If modified, the author asks for a re-review. The reviewer checks the modified files and may approve the change if he/she is satisfied. If not, the reviewer may ask for further changes. This process may repeat for several iterations until the reviewer(s) approve the change or it is abandoned. 

 \subsection{Regression analysis}
Regression analysis is a powerful statistical approach that helps to analyze how one or multiple independent variables influence one dependent variable \cite{foley2018regression}. There are two primary uses of regression-- (i) predictive analysis and (ii) inferential analysis \cite{allison2014prediction}. In predictive analysis, the goal is to develop a formula to predict the value of a dependent variable based on the values of one or more independent variables. In comparison, the goal of an inferential analysis is to determine whether a particular independent variable has any impact on the dependent variable and the magnitude of that impact if it exists. The inferential analysis differs from the predictive analysis preliminary based on two key factors. First, \emph{multicollinearity}: if two or more independent variables are highly correlated to each other, considering all of those correlated variables together can produce an over-fitting problem in inferential analysis. However, in predictive analysis, multicollinearity is not an issue. Second, $R^2$: $R^2$ represents the goodness of fit of a regression model \cite{helland1987interpretation}. Higher \emph{$R^2$} is important; however, it is more important in predictive analysis. In an inferential analysis, even with a small $R^2$, the regression model can provide useful insights regarding the relationships between the independent and the dependent variables \cite{allison2014prediction}. \revision{We have used inferential regression analysis in this work.}

\section{Research Questions}
 \label{sec:reseach-questions}
 The following subsections introduce the two research questions guiding this research.
 
\subsection{RQ1: What makes a code review useful to OSS developers?}

These insights may help OSS participants practice code reviews to improve CR effectiveness. However, `usefulness' as a complex phenomenon may have multiple associated facets. 
 We divide this question into two sub-questions to investigate technical and non-technical facets of CR usefulness. \revision{ Our first sub-question (i.e., RQ1.A) aims to find out the open-ended opinions of OSS developers about CR usefulness to identify whether the CR usefulness criteria for OSS developers are similar to the ones considered by industry participants~\cite{bosu2015characteristics,bosu2016process,bacchelli2013expectations}. More specifically, we investigate how primary (e.g., identification of defects, code improvement opportunities, or alternative solution approaches) and secondary CR usefulness criteria (e.g., improving code maintainability, facilitating knowledge sharing, or helping relationship formation among teammates.) reported by prior studies rank among OSS participants.}

\vspace{4pt}
  \emph{RQ1.A: What are the open-ended opinions of OSS developers regarding code review usefulness?}
\vspace{4pt}

 As we aim to rank CR comments based on the degree of usefulness, some of the secondary CR usefulness criteria, such as relationship formation and knowledge sharing, are \revision{difficult to rate by an independent evaluator within the limited context and timeframe of a user study.   Therefore, we narrow down our focus into a facet that a study participant can independently determine}, i.e., the type of comment authored in a CR. Hence, we ask: 

\vspace{4pt}
   \revision{\emph{RQ1.B: How do OSS developers rank various categories of code review comments based on their degrees of usefulness?}}

\subsection{RQ2: Which contextual and participant characteristics are associated with CR usefulness?}
\revision{We term attributes of a CR context (i.e., where a CR occurred) as contextual factors. For example, characteristics of changes under review (e.g., code churn and number of files) belong to contextual factors. On the other hand, attributes of the CR participants (e.g., author's /reviewers' experience and prior history with the files under review) belong to participant characteristics. 
We focus only on contextual and participant-related factors since we aim to identify actionable recommendations for review preparation and reviewer selection to maximize the likelihood of receiving useful feedback. Table~\ref{table:usefulness-factors} provides a list of contextual and participant factors selected based on prior studies.} 
We divide this question into two sub-questions. Our first sub-question aims to identify how various contextual and participant factors may contribute to maximizing CR usefulness.

\vspace{4pt}
 \revision{\emph{RQ2.A: How are various contextual and participant characteristics associated with the  degree of usefulness achieved in a code review?}}
\vspace{4pt}

Our second sub-question aims to maximize primary CR usefulness criteria, such as identifying functional defects. Therefore, we investigate how various factors are associated with these primary usefulness criteria instead of secondary ones.  

\vspace{4pt}
 \emph{RQ2.B: How are various contextual and participant factors associated with identifying functional defects? }

\section{Research Methodology}
\label{sec:methods}
This section details our project selection, data collection, survey design, and qualitative and quantitative analyses approach.

\subsection{Project selection}
\revision{We selected the OpenDev community for the survey and selected the OpenDev Nova project for  our quantitative analysis for the following reasons:}
\begin{itemize}
    \item OpenDev is one of the largest OSS development communities with more than 15K contributors, and OpenDev developers practice tool-based CRs \cite{bosu2016process}.
    \item The OpenDev family includes 52 different projects as of 2023~\cite{openstack-project-list}, which includes Nova, Neutron, Heat, and Swift.
    \item OpenDev community includes approximately 110k persons from  more than 700 organizations spanning 182 countries worldwide \cite{openstack-report}.    
    \item OpenDev Nova is one of the most active projects in the OpenDev community \cite{hirao2020code}, and it has been the subject of prior studies investigating CRs~\cite{han2021understanding,zanaty2018empirical}.
\end{itemize}

\subsection{Data Mining} 
The CR repository of the OpenDev community is managed by Gerrit\footnote{https://www.gerritcodereview.com/}. We use a Java-based Gerrit Miner tool to access Gerrits' REST API to mine all the publicly available CRs from the OpenDev's  repository\footnote{https://review.opendev.org} and store those in a MySQL database. Our dataset includes a total of 795,226  completed (i.e., either `Merged' or `Abandoned') CRs spanning July 2011 to March 2022.

\subsection{Classification Rubric}
\begin{figure}
	\centering  \includegraphics[width=\linewidth,trim={3cm 4cm 5cm 0}]{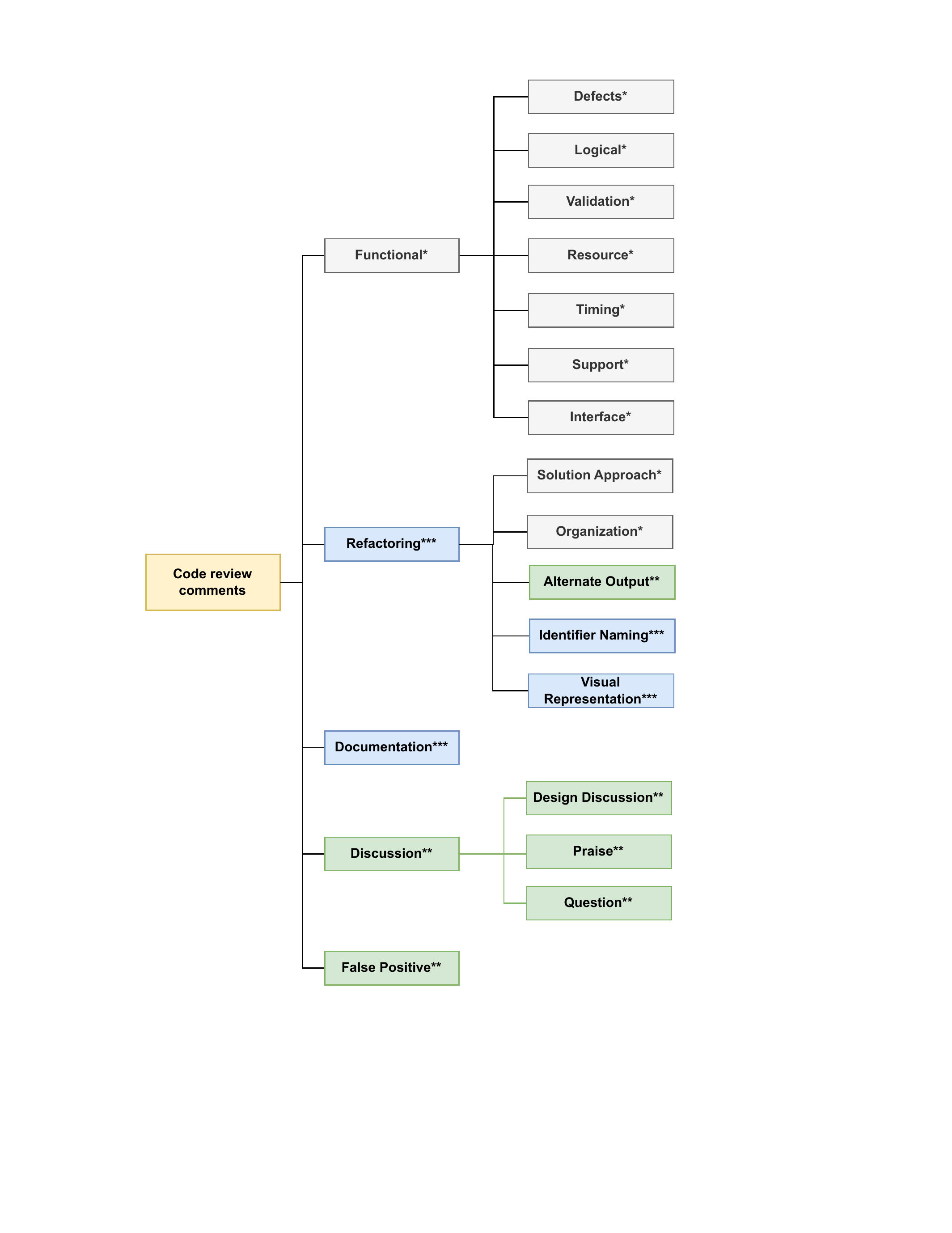}
	\caption{Adopted code review comment classification scheme. \revision{* $\rightarrow$ components adopted directly from Beller \textit{et} al., ** $\rightarrow$ newly added categories, and *** $\rightarrow$ modification of Beller \textit{et} al. scheme.}}
	\label{fig:classification_scheme}		
\end{figure}

We have identified five prior studies that have presented classifications of issues identified during CRs~\cite{beller2014modern,bosu2015characteristics,mantyla2008types,Panichella2020empirical,hasan2021using}. We started with the classification proposed by Beller et al.\cite{beller2014modern}. In the proposed classification scheme, Beller \textit{et} al. classified CR changes into two higher-level categories: `Functional' and `Evolvability'. They divided `Functional' and `Evolvability' into six and three sub-categories, respectively. Two of the `Evolvability' sub-categories (`Structure', and `Documentation'), were further divided to form two lower-level sub-categories from each.

While we use their scheme as the basis for our proposed one, we made the following three modifications to fit our study context better.
\begin{enumerate}
    \item Beller \textit{et} al.'s scheme is targeted towards CR changes~\cite{beller2014modern}, whereas we focus on CR comments. Prior studies have found a non-trivial number of CR comments that do not trigger code changes~\cite{bosu2015characteristics,hasan2021using}. For example, Bosu \textit{et} al. found four such categories:  i) `Praise', ii) `Questions', and iii) `Design discussion'~\cite{bosu2015characteristics}. Later Hasan \textit{et} al. also found those categories of CR comments~\cite{bosu2015characteristics}. Therefore, we introduce a new category named `Discussion' that includes these three sub-categories.
    
    \item \revision{ A `False positive' is a CR comment where the reviewer's concern is invalid.} Beller \textit{et} al. excluded false positives~\cite{beller2014modern}, since those do not induce changes. We include a new category named `False positive' to include those, as we target CR comments instead of changes. 
    
    \item Instead of four levels in Beller \textit{et} al.'s schema~\cite{beller2014modern}, we simply our schema into three levels by excluding the `Evolvability' category and putting its subcategories directly under the top level. 
    
    \item We rename the `Structure' subcategory from Beller \textit{et} al. into refactoring, since types of changes belonging to fall under Martin Fowler's refactoring catalog~\cite{fowler2012refactoring}. 
    We add the `Alternate output', another sub-category discovered by Bosu et al.~\cite{bosu2015characteristics} under this group as well.
    Finally, we also include `Visual representation' under `Refactoring', as those were included among refactorings by prior studies~\cite{zaidman2011studying,hassan2003studying}.   
    
\end{enumerate}

 Figure~\ref{fig:classification_scheme} shows our CR classification scheme. Components in \revision{one star (*)} are taken from Beller \textit{et} al~\cite{beller2014modern} without modification. Components in \revision{two stars (**)} and \revision{three stars (***)} are our additions and modifications, respectively. Table~\ref{table:rubric} provides a brief description of each of the CR comment categories from this scheme.

\begin{table*}
\caption{Description of the categories from our code review comment classification scheme}
\centering \label{table:rubric}
\begin{tabular}{ |p{2.3cm} |p{3cm} | p{10cm}| } 
\hline
\textbf{Group} & \textbf{Category} & \textbf{Description}   \\ 
\hline
\multirow{7}{*}{\textbf{Functional}}
& Functional defect & Functionality is missing or implemented incorrectly and such defects often require additional code or larger modifications to the existing solution. \\ \cline{2-3}
& Logical & Control flow, comparison related, and logical errors. \\ \cline{2-3}
& Validation & Validation mistakes or mistakes made when detecting an invalid value are of this class. Any kind of user data sanitization-related comments are in this category, too. \\ \cline{2-3}
& Resource & Resource (variables, memory, files, database) initialization, manipulation, and release.  \\ \cline{2-3}
& Timing & Potential issues due to incorrect thread synchronization. \\ \cline{2-3}
& Support & Issues related to support systems and libraries or their configurations.  \\ \cline{2-3}
& Interface & Mistakes when interacting with other parts of the software such as - existing code library, hardware device, database or operating system. 
\\ 
\hline
\multirow{5}{*}{\textbf{Refactoring}}
& Solution approach & Suggestions to adopt an alternate algorithm or data structure. \\ \cline{2-3}
& Alternate Output  & Comments that suggest modifying the error message, toast message, alert, or change what is returned by a function. \\ \cline{2-3}
& Organization of the code & Refactoring suggestions such as those included in Martin Fowlers's catalog~\cite{fowler2012refactoring}. \\ \cline{2-3}
& Naming Convention & Violations of identifier naming conventions. \\ \cline{2-3}
& Visual Representation & Whitespace, blank lines, code rearrangements, and indentation-related comments. \\  \cline{2-3}
 \hline

 \multirow{1}{*}{\textbf{ Documentation}}& 
 Documentation & Suggestions to add /modify comments or documentation to aid code comprehension. \\ 
\hline

\multirow{3}{*}{\textbf{Discussion}} & Design discussion & Discussions on design direction, design pattern, and software architecture. \\ \cline{2-3} 

 & Question & Questions to understand the design or implementation choices. \\ \cline{2-3}
 & Praise & Complement for a code.\\ \hline

\textbf{False positive} & False positive & If a review comment raises an invalid bug or concern. \\ 
\hline

\textbf{Others}  & Other & Comments not belonging to any of the above categories.\\
\hline

\end{tabular}
\end{table*}

\subsection{Manual Labeling}
 For this phase, we randomly selected  2,500 CR comments from the OpenStack Nova project. We chose OpenDev Nova has the highest number of distinct contributors as well as the highest number of completed CRs within the OpenDev community. According to an investigation of our dataset, \revision{during the period of July 2011 to March 2022, Nova has 38,775 either completed or abandoned CRs. The second highest is the Neutron, which has 23,456 either completed or abandoned CRs during the above-mentioned period. Our collected dataset of the Nova project contains a total of 300,304 code review comments from July 2011 to March 2022.} We selected 2,500 as the sample size to achieve a 95\% confidence interval with a 2\% margin of error~\cite{yamane1973statistics}.
Two of the authors independently categorized each of the selected comments into one of the eighteen categories from Table~\ref{table:rubric} after reading the entire discussion thread associated with a comment and also inspecting its associated code context.  Each author also labeled each comment as either `useful' or `not useful'. \revision{
For this labeling, we adopt the `usefulness' labeling rubric proposed by Bosu \emph{et} al.~\cite{bosu2015characteristics}. Two subsequent CR usefulness studies~\cite{rahman2017predicting,han2021understanding} have also used this rubric, which is as follows:}
\begin{itemize}
    \item \textit{Useful}-- A code review comment is considered `useful':
    
    \begin{enumerate} [noitemsep,topsep=0pt]
    \item  if the code author explicitly acknowledges the identified issue (e.g., ``Good catch'').
    \item If the code author implicitly acknowledges the identified issue by making the suggested changes.
    \item If the author explicitly defers the identified issue to a future change.
\end{enumerate}

\item \textit{Not useful}-- A comment is considered `not useful':
\begin{enumerate} [noitemsep,topsep=0pt]
    \item  if the code author explicitly states the issue as invalid.
    \item  if the code author neither makes the suggested changes nor acknowledges deferring it to a future change. 
\end{enumerate}
\end{itemize}

We computed the level of inter-rater reliability of this multi-label manual categorization process (assigning comments into 18 categories) using Cohen's kappa ($\kappa$)~\cite{cohen1960coefficient} which was measured as 0.68 (`a substantial agreement'\footnote{Kappa ($\kappa$) scores are commonly interpreted as: 0.01–0.20 as `none to slight,' 0.21–0.40 as `fair,' 0.41– 0.60 as `moderate,' 0.61–0.80 as `substantial,' and 0.81–1.00 as `almost perfect agreement'~\cite{landis1977application}.}). Cohen's kappa value for the `useful'/`not useful' categorization is 0.84.

\begin{table*}
	\centering
    \caption{Questions included in our survey}    
	\label{table:survey-questions}
\begin{tabular}{p{0.5cm} p{0.5cm} p{9cm} p{5cm}}
\hline
\textbf{\#} & \textbf{RQ\#} & \textbf{Question Text} & \textbf{Answer Choices} \\
\hline 
Q1 & D & What is your highest level of education? & [ High school, Bachelors, Masters, Ph.D., Other ] \\
Q2 & D & How many years of software development experience do you have? & [ Numeric input] \\
Q3 & D & How many years have you been practicing tool-based code reviews? & [ Numeric input] \\
Q4 & D & How many hours per week, on average, do you spend reviewing other contributors' code? & [Numeric input ] \\
Q5 & RQ1 & In your opinion, what makes a code review useful? & [Free form text] \\
Q6 & D & Approximately, what proportion of code reviews in your project, do you feel are useful (i.e., fits the above definition)? & [ Numerical slider from 0\%-100\%] \\
Q7 & RQ1 & Rate the following categories of review comments based on their perceived usefulness to you. Choices are 16 categories from Table \ref{table:rubric} excluding the `False positive' and `Others' categories.  & Matrix table~\cite{qualtrics-questions} with [1-5 Star Rating]\\
Q8-Q39 & RQ1 & How useful do you find the above code review on a scale of 1 to 10 if you were the author? & Each question showed a screenshot such as Figure~\ref{fig:examples} [1-10 Star Rating] \\
\hline 
\end{tabular}

\end{table*}
\begin{figure*}
     \centering
     \begin{subfigure}[b]{0.75\textwidth}
         \centering
         \includegraphics[width=\columnwidth]{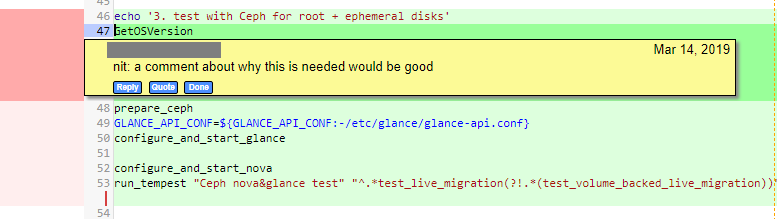}%
         \caption{Documentation}
         \label{fig:sample1}
     \end{subfigure}
    ~ \hfill
     \begin{subfigure}[b]{.75\textwidth}
         \centering
         \includegraphics[width=\columnwidth]{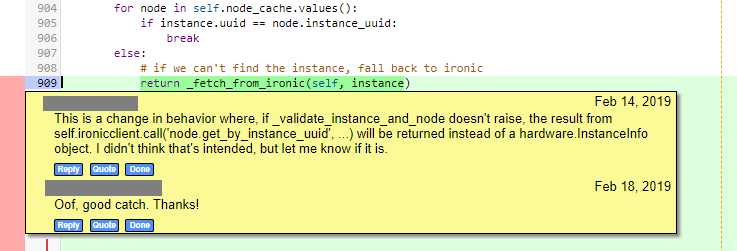}
         \caption{Validation}
         \label{fig:sample-2}
     \end{subfigure}
        \caption{Examples of code review comments included in our survey}
        \label{fig:examples}
\end{figure*}

\subsection{Survey Design for RQ1}
\label{sec:survey-design}

To better understand what makes a CR useful to OSS developers (RQ1), we designed an online survey for persons who have first-hand knowledge of participating OSS CRs \revision{(i.e., actively participated in CRs by providing suggestions)}. 

\revision{\subsubsection{Questionnaire preparation}}

Since RQ1.A aims to find out the open-ended opinions of OSS developers about CR usefulness. Our survey asks, `What makes a code review useful (Q5 in Table~\ref{table:survey-questions})?'.
RQ1.B aims to understand how useful each kind of code review comment listed in Table~\ref{table:rubric} is.
On this goal, we include two categories of questions in the survey. First, we ask the respondents to manually rate each CR category on a five-point \revision{Likert} scale, only based on a short definition of that category (Q7). Second, we ask the respondents to rate sample CR comments on a 10-point scale, by only showing screenshots and without revealing the actual categories those comments belong to (Q8 to Q39), \revision{in total, there were 32 CR comment samples}.  
To identify sample CR comments for this 10-point scale rating, we identified multiple ideal examples for each CR comment category from our manually labeled dataset. We consider an example ideal if its category can be distinguished unambiguously by reading the comment and its context (i.e., screenshot). To be an ideal example, a CR comment: i) must clearly state what needs to be changed or what the comment pertains to, and ii) its surrounding code context must include the relevant code snippet.

However, we excluded the `False positive' category as those are not useful (i.e., 0). We also excluded `Others,' as \revision{we could not find} a representative sample for that category. With the remaining 16 CR comment categories, two of the authors independently investigated our labeled dataset. They identified a total of 64 candidate examples, where exactly four examples belonged to each CR comment category. 
We merged these two lists, assigning a score of `2' to examples belonging to both lists and assigning `1' to others.  
Next, we had a discussion section to pick two of the most representative examples for each CR comment category, where examples with a score =2, got higher priorities.  At the end of this process,  we selected a total of 32 examples. We took Gerrit screenshots of the selected examples with surrounding code contexts. Figure~\ref{fig:examples} shows two of the examples included in our survey.  Table~\ref{table:survey-questions} lists the questions included in our survey. We have also included all the questions and 32 selected examples in our supplementary material.

\subsubsection{Pilot Survey}
For expert reviews, we sent the survey questionnaire to two renowned Software Engineering (SE) researchers with expertise in code reviews. According to their feedback, we made several edits for structural and linguistic modifications. Finally, we piloted the survey with \revision{three} SE  graduate students for validation. \revision{The goal of piloting the survey is to identify any visible mistakes and to get any modification suggestions for improving the survey questions.} We got the final survey questions, consent form, participant selection strategy, solicitation email, and data management protocol reviewed and approved by our university's Institutional Review Board (IRB).

\subsection{Data Collection}
To solicit the opinions of participants with adequate CR experience, we sent our survey only to developers who have submitted at least 5 code changes for review. Since all the code snippets are written in Python, we also limited our selection to OpenDev projects with Python listed as the primary language. We identified a total of 4,188 OpenDev developers satisfying these criteria.  

We sent personalized emails, composed according to the guidelines provided by our IRB, to each of the 4,188 developers with the link to the survey hosted on Qualtrics~\cite{snow2013qualtrics}. Since we obtained the email addresses by mining Gerrit, there are ethical considerations~\cite{baltes2016worse} for sending such invitations. Therefore, each of our invitations:  i)  indicated the OSS project that we mined to obtain his/her email address, ii) indicated if the participant would get any more emails from us,  iii) provided a link to opt-out, and iv) apologized for the inconvenience.
Excluding the 1,010 \revision{undelivered ones (i.e., email address no longer valid)}, we consider 3,178 emails as delivered. We obtained a total of 238 responses, a response rate of 7.49\%.

\subsection{Survey data analysis}
 We checked the quality of the recorded responses and excluded 78 cases where the respondent did not answer at least one survey question that is tied to our research questions (i.e., left the survey after completing only the demographic section). So, we conducted our analysis on 160 responses.
 For the one qualitative question (i.e., Q5 in Table~\ref{table:survey-questions}), we adopted a systematic data analysis process.  Two of the authors independently went through all the responses to extract the factors of useful CRs mentioned by our respondents. Then we had a discussion session to cross-validate and create a finalized list of factors. With this list, two of the authors independently labeled each response to one or more codes. After completion, we compared the labels to identify conflicts. Finally, we had a discussion session to resolve those conflicting labels.  Inter-rater reliability based on Cohen's kappa $\kappa$ for this labeling was 0.56 (i.e., `moderate'). The low $\kappa$ value is because a single response might mention one or multiple CR usefulness criteria, which creates theoretically exponential possibilities.

\begin{table*}
	\centering
    \caption{Respondents' demographics}    
	\label{table:demographics}
\begin{tabular}{|p{7cm}|C{1.2cm}|C{1.5cm}|L{2cm}|R{2.3cm}|}
\hline
\textbf{Question}                                                                         & \textbf{Mean} & \textbf{Median} & \textbf{Categorical description}                                                                                                              & \textbf{\% of respondents}                                                      \\ \hline
Q1. What is your highest level of education?                                              & ---           & ---             & \begin{tabular}[c]{@{}l@{}}High school\\ Bachelors\\ Masters\\ Ph.D.\\ Other\end{tabular}                                                     & \begin{tabular}[c]{@{}c@{}}36.6\%\\ 50.7\%\\ 5.4\%\\ 4.4\%\\ 2.9\%\end{tabular} \\ \hline
Q2. How many years of software development experience do you have?                        & 10.14 years   & 10 years        & \begin{tabular}[c]{@{}l@{}}$<$ 2 years\\ 2-5 years\\ 6-10 years\\ $>$ 10 years\end{tabular}     & \begin{tabular}[c]{@{}c@{}}1.5\%\\ 19.5\%\\ 33.7\%\\ 45.3\%\end{tabular}        \\ \hline
Q3. How many years have you been practicing tool-based code reviews?                      & 5.43 years    & 5 years         & \begin{tabular}[c]{@{}l@{}}$<$ 2 years \\ 2-5 years \\ 6-10 years \\ $>$ 10 years\end{tabular}  & \begin{tabular}[c]{@{}c@{}}10.2\%\\ 44.9\%\\ 28.8\%\\ 16.1\%\end{tabular}       \\ \hline
Q4. How many hours per week, on average, do you spend reviewing other contributors' code? & 5.10 hours    & 4 hours         & \begin{tabular}[c]{@{}l@{}}$<$ 2 hours \\ 2-5 hours \\ 6-10 hours\\ $>$ 10 hours\end{tabular} & \begin{tabular}[c]{@{}c@{}}23.9\%\\ 42.9\%\\ 15.1\%\\ 18.1\%\end{tabular}       \\ \hline
\end{tabular}
    
\end{table*}

Table~\ref{table:demographics} provides an overview of the demographics of our survey respondents. Our respondents include mostly experienced software developers, with 79\% having more than five years of software development experience. The respondents also have significant code review experiences, as almost 90\% have been practicing tool-based code reviews for more than two years. On average, our respondents spend 5 hours per week in code review, which is on par with the numbers reported in prior studies~\cite{bosu2016process,hasan2021using}. Based on these characteristics, we believe that our respondents are eligible to provide valuable opinions to answer our research questions.

\section{Results: (RQ1) What makes a code review useful to OSS developers?} 
\label{sec:results-rq1}
The following subsections describe the results of our first research question based on our survey responses.

\begin{table*}
	\centering
    \caption{Codes that emerged from open-ended survey question and category that we assigned each code to}
	\label{table:codes}
\begin{tabular}{p{6cm}|l}
\hline
  \textbf{Assigned category} (\% respondents)   & \textbf{Theme (\# of respondents)}               \\ \hline
  \multirow{4}{*}{Find defects (44.4\% )}  & Defect finding (126)   \\ \cline{2-2}
   &          Security issues (11)        \\ \cline{2-2}
   &          Fulfill requirements (11)    \\ \cline{2-2}
   &          NIT/typos (9)              \\ \cline{1-2} 
   \multirow{3}{*}{Improve code quality (42.5\% )}  &  Code optimization (48)  \\ \cline{2-2}
  &      Alternative implementation (35)           \\ \cline{2-2}
  &           Improve documentation (21)   \\ \cline{1-2} 
  \multirow{5}{4cm}{Uses appropriate language (28.1\%)}    &  Constructive criticism (29)  \\ \cline{2-2} 
        &      Respectful (15) \\ \cline{2-2} 
                 & Concrete example (13)     \\ \cline{2-2}
  & Understandable (12)                \\ \hline

  \multirow{3}{*}{Maintainability (28.1\%)}   & Code comprehensibility  (23)           \\ \cline{2-2}
  &           Maintain design (17)  \\ \cline{2-2}
  &          Reduce maintenance cost (6)         \\ \cline{1-2} 
   \multirow{3}{*}{Facilitates knowledge sharing (25.0\%)}   & Knowledge transfer (35)     \\ \cline{2-2}
  &           Mentoring (3)                \\ \cline{2-2}
  &           Project awareness (10)          \\ \cline{1-2} 
    \multirow{3}{*}{Facilitates better design (24.4\% )}  &  Design suggestion (27)      \\ \cline{2-2}
  &           Consider wider context (13)             \\ \cline{2-2}
  &           Questions leading to better design (12)  \\ \cline{1-2} 
  Appreciate good work (3.1\%)   &    Appreciate good work (5)    \\ \cline{1-2} 
  Helps community building (3.1\%)   & Helps community building (5)         \\ \hline
   Timely feedback (1.8\%)   & Timely feedback (3)         \\ \hline
\end{tabular}
\end{table*}

\subsection{RQ1.A: What are the open-ended opinions of OSS developers regarding code review usefulness?}
Based on our respondents' opinions, we identified 23 different criteria (codes) to judge a CR as useful. Posthoc, we aggregated those codes into 9 higher-level groups. Table ~\ref{table:codes} shows those nine groups, 23 codes, and the percentage of respondents mentioning the code. The sum of the percentages exceeds 100\% since respondents could mention multiple codes in their responses. Unsurprisingly, `Finds defects' tops this list, followed by `Improve code quality'. However, we notice a strong emphasis on non-technical aspects, with one-third of respondents judging CR's usefulness based on linguistic characteristics. We also ran Chi-Square ($\chi^2$) tests to check whether these criteria differed based on education level, software development experience, or code review experience. We found significant differences ($\chi^2 =13.34,  p=0.012$)\footnote{ $p-values$ are adjusted using Benjamini and Hochberg corrections~\cite{benjamini1995controlling} due to multiple comparisons.} only based on software development experience as veteran developers with more than ten years of experience were more likely to use the `finds defect' criteria.

The following subsections detail those nine CR usefulness criteria based on our respondents' opinions. We also include verbatim excerpts to illustrate their opinions further. Each excerpt is associated with a numeric value that indicates a unique identifier for the respondent who expressed that opinion. For example, [\#40] indicates a response from the 40$^{th}$ respondent.

\subsubsection{{Finds defects}}
Most of the respondents (44.4\%) consider a code review as useful if it points out a defect. Identification of any functional defects (i.e., the functional group from Table~\ref{table:rubric}) is useful and is one of the primary expectations of code authors~\cite{bacchelli2013expectations}.
\surveyquote{Points to errors or issues, that the author didn't see during writing code. [\#40]
}

Prior research found evidence of security defects identified during code reviews~\cite{Bosu-et-al-FSE:2014,paul2021security}. Although the identification of security defects is rare, they are extremely useful.
\surveyquote{Any code review that aims to improve the code is useful. Sometimes the improvement is drastic and critical, such as pointing out an error that would lead to a crash, or a security issue. [\#116]
}

Code reviews can also serve as a validation step as reviewers may provide feedback on requirement fulfillment.
\surveyquote{Proper scoping of the reviews:- Architectural reviews, alignment with requirements ... [\#52] }

\subsubsection{{Improves code quality}}
Any suggestions to improve code quality are useful.  A large number of our respondents appreciate suggestions to achieve better quality through an alternate solution approach.

\surveyquote{Clear, constructive discussion of the code, ..., suggestions from reviewers for alternative implementations if there are areas the code could be improved or does not meet criteria. [\#103]
}

Code optimization is another quality improvement area, as reviewers might suggest possible code reuse opportunities and code optimization techniques.

\surveyquote{Optimization of code e.g. newcomers may not know full design so they may implement an already existing function which could be found in code review. [\#206] }

Documentations to aid program comprehension are crucial for long-term maintenance. Pointing out mistakes in documentation or documentation improvement suggestions is useful.
\surveyquote{ Finding structural issues/code comments /documentation that would make the code hard to maintain going forward. [\#130]
}

\subsubsection{{Uses appropriate language}}
To be useful, a CR comment needs to be authored in an appropriate language as one of our respondents mentioned, \surveyquote{.. it's not only about what was written, but also how it was written. [\#56] }

Harsh critiques may demotivate an author and are less likely to be useful. Moreover, reviews must be concise, understandable, and explanatory so that authors can easily understand their mistakes and make necessary changes.

\surveyquote{I typically write code reviews in the form of `appreciate - suggest a change or point out the mistake - appreciate again' This seems really useful to me because it does not undermine the developer's confidence and at the same time tells them about how to go about enhancing or correcting their approach. [\#221] }

Sometimes examples illustrating potential problems are extremely useful to help authors understand problems with their proposed solutions. 
\surveyquote{...reviewers must clearly state their concerns and opinions and where applicable point to specific code lines, documentation, mailing list threads, or other discussions that are relevant to the matter at hand or provide examples. [\#10] }

Even if a reviewer rejects a change, it is important to do so in a constructive way.
\surveyquote{If negative, provides constructive feedback, the more definitive the better. [\#101] }


\subsubsection{{Improves `maintainability'}}

Code reviews checking for violations of project design constraints are also useful
\surveyquote{1. what are the problems about the code? 2. Does the code align with the project's blueprint?  [\#141] }

Code that is not readable is difficult to maintain. Future contributors will face difficulty to modify the less readable code. Suggestions to improve readability from a different person's perspective are useful.
\surveyquote{...pointing and discussing better ways of achieving the same result. by better ways I mean: code readability, performance, and maintainability. [\#215] }

\subsubsection{{Facilitates knowledge sharing}}
Prior studies suggest knowledge sharing as one of the most important benefits of code reviews~\cite{bacchelli2013expectations,bosu2016process}. Code review is a bidirectional knowledge transfer process, the code owner learns from the reviewers' comments and the reviewer also learns from the code owner. Reviews facilitating such knowledge transfer are useful.
\surveyquote{Knowledge transfer.  It lets other members of the community share their experiences with others.  Sometimes the more experienced review participant knows a better way to do things. Sometimes the more experienced review participant is aware of past design decisions or design constraints the other party is not aware of. [\#8] }

A useful review facilitates the newcomers to learn code architecture, team culture, and coding conventions.
\surveyquote{Friendly relations with the core developers, other participating developers, and mentorship. [\#166] }

\subsubsection{{Facilitates a better design}}

Reviewers provide useful suggestions on design direction and design patterns which makes software products better architectured. 
\surveyquote{...lead to a discussion about potential better solutions (for bug fixes) or future design (for implementing new features). [\#121] }

Code review is the platform for reviewers to identify errors and for the code owner to defend their work or make it correct. Such discussions help increase the breadth of knowledge of both reviewers and code owners.
\surveyquote{Providing the wider context (e.g. interactions with modules or systems the code author did not consider). [\#1] }

Asking critical questions about software architecture leads to a better design of the product. Moreover, when the reviewer asks questions about some change, that means the code is not sufficiently self-explanatory. So from the readability perspective, there is an opportunity for improvement.
\surveyquote{Highlighting issues in the code or asking relevant questions to the topic which could lead to improvements. [\#111] }

\subsubsection{{Appreciates good work}}
Positive encouragement is essential for inspiring new contributors to do quality work and increasing their confidence. Many respondents find such reviews useful.
\surveyquote{... code contributor gets more confident in the code that has been appreciated  and is encouraged to continue contributing. [\#41]}

\subsubsection{{Timely feedback}}
As review delays can frustrate authors, useful feedback should be timely and not create confusion.
\surveyquote{...Ideally, all concerns should be expressed at once, avoiding back and forth communication between the reviewer and the code author,  which can significantly delay the patch approval. [\#88] }

\subsubsection{{Helps community building}}
Constructive criticism creates openness and better programming habits within a community and therefore is useful.
\surveyquote{Build team culture for better programming habits, confidence in the artifacts, openness in communication. [\#6] }

 \begin{boxedtext}
\textbf{Key takeaways:} \emph{According to OpenDev developers, a CR's  usefulness is dictated not only by its technical contributions such as defect finding or quality improvement tips but also by linguistic and process aspects such as comprehensibility, tone, and timeliness.}
\end{boxedtext}

\begin{table*}
	\centering
    \caption{Ranking of code review comments based on usefulness rating}   
	\label{table:usefulness-ranking}
	\resizebox{\linewidth}{!} {
\begin{tabular}{|p{1.9cm}|c|r|r|c|r|r|R{1.6cm}|R{1.2cm}|}
\hline
\multirow{2}{1.7cm}{\textbf{Comment category}}  &\multicolumn{3}{c|} {  \textbf{User perception}} & \multicolumn{3}{c|}{  \textbf{Sampled user rating}} & \multirow{2}{1.6cm}{ \textbf{Ratio of comments}} &  \multirow{2}{1.2cm}{\textbf{Avg. Rating}}  \\  \hhline{~------~~}
&  \textbf{Distribution}& \textbf{Avg ($\mu$)}. & \textbf{SD ($\sigma$)} & \textbf{Distribution} & \textbf{Avg. ($\mu$)} & \textbf{SD ($\sigma$)} &  & \\
\hline
Functional Defect & \tickerRatingTeal{0}{4.5}{7.7}{21.2}{66.7} &  4.50 & 0.82 & \tickerRating{0}{2.6}{12.6}{41.1}{43.7}& 4.26 & 0.89 & 0.48\% & 4.38\\ \hline

Validation & \tickerRatingTeal{3.9}{5.2}{27.1}{34.2}{29.7} & 3.81 & 1.05 & \tickerRating{0}{0.7}{8.0}{31.4}{59.9}& 4.50 & 0.73 & 3.68\% & 4.16\\ \hline

Logical & \tickerRatingTeal{0}{3.2}{12.8}{13.5}{70.5} & 4.51 & 0.84 & \tickerRating{1.9}{11.5}{23.1}{40.4}{23.1}& 3.71 & 1.16 & 2.28\% & 4.11\\ \hline

Interface & \tickerRatingTeal{1.9}{3.9}{14.8}{32.3}{47.1} & 4.19 & 0.96 & \tickerRating{0}{4.8}{20.7}{44.1}{30.3}& 4 & 0.96 & 1.60\% & 4.10\\ \hline

Solution Approach & \tickerRatingTeal{1.3}{5.1}{19.2}{29.5}{44.9} & 4.12 & 0.98 & \tickerRating{0}{4.5}{27.6}{43.3}{24.6}& 3.88 & 0.92 & 8.48\% & 4.00\\ \hline

Question & \tickerRatingTeal{1.9}{2.6}{17.9}{41.0}{36.5} & 4.08 & 0.91 & \tickerRating{0.7}{2.2}{24.6}{52.2}{20.1}& 3.89 & 0.93 & 13.56\% & 3.99\\ \hline

Design Discussion & \tickerRatingTeal{4.5}{5.8}{15.5}{26.5}{47.7} & 4.07 & 1.13 & \tickerRating{0.7}{11.6}{29.5}{37.0}{21.2}& 3.66 & 1.21 & 3.64\% & 3.87\\ \hline

Resource & \tickerRatingTeal{4.5}{8.3}{27.6}{33.3}{26.3} & 3.69 & 1.09 & \tickerRating{0}{3.8}{19.1}{54.2}{22.9}& 3.96 & 1.20 & 1.48\% & 3.83\\ \hline

Documentation & \tickerRatingTeal{2.6}{5.1}{28.2}{34.6}{29.5} & 3.83 & 1.0 &  \tickerRating{0.7}{13.8}{28.3}{35.9}{21.4}& 3.63 & 1.17 & 33.32\% & 3.73\\ \hline

Organization of Code & \tickerRatingTeal{5.1}{10.3}{24.4}{37.8}{22.4} & 3.62 & 1.10 & \tickerRating{0.7}{6.6}{29.2}{45.3}{18.2}& 3.74 & 1.0 & 7.68\% & 3.68\\ \hline

Alternate Output & \tickerRatingTeal{4.5}{14.1}{35.9}{28.8}{16.7} & 3.39 & 1.06 & \tickerRating{1.3}{4.7}{23.3}{47.3}{23.3}& 3.87 & 0.93 & 2.56\% & 3.63\\ \hline

Timing & \tickerRatingTeal{6.5}{9.1}{29.2}{26.6}{28.6} & 3.62 & 1.18 & \tickerRating{3.8}{15.2}{30.3}{41.7}{9.1}& 3.37 & 1.12 & 0.16\% & 3.5\\ \hline

Naming Convention & \tickerRatingTeal{9.0}{19.9}{34.6}{20.5}{15.4} & 3.11 & 1.20 & \tickerRating{2.8}{8.3}{27.6}{33.8}{27.6}& 3.75 & 1.10 & 3.52\% & 3.43\\ \hline 

Praise & \tickerRatingTeal{18.1}{16.1}{29.7}{21.3}{14.8} & 2.99 & 1.30 & \tickerRating{12.6}{19.3}{30.4}{24.4}{13.3}& 3.07 & 1.35 & 4.20\% & 3.03\\ \hline

Visual Representation & \tickerRatingTeal{18.5}{15.9}{23.6}{16.6}{24.8} & 3.12 & 1.45 & \tickerRating{25.8}{18.9}{25.8}{17.4}{12.1}& 2.71 & 1.50 & 3.92\% & 2.92\\ \hline 

\end{tabular} 
}
\end{table*}

\subsection{RQ1.B: How do OSS developers rank various categories of code review comments based on their degrees of usefulness?} 

In response to Q7 in Table~\ref{table:survey-questions}, our respondents rated 16 CR comment categories on a five-point scale. The three columns under the \emph{User perception} header in Table~\ref{table:usefulness-ranking} show the distributions of the ratings provided by our respondents, the average rating for each comment category, and the standard deviation of the assigned ratings. 

Q8-Q39 asked the respondents to rate 32 selected example code reviews on a ten-point scale based on their usefulness (i.e., how useful they find a code review on a scale of 1 to 10 if they were the author). Since we included two examples for each category, a respondent's \emph{Sampled user rating} for a comment category was computed by taking the average of their two scores for the examples of that particular category. For comparison against the \emph{User perception} score of a category, we divided the average \emph{Sampled user rating} by 2.  The three columns under the \emph{Sampled user rating} header in Table~\ref{table:usefulness-ranking} show the distributions of the ratings assigned to sample code reviews from each category. The  \emph{Ratio of comments} column shows the percentage of review comments belonging to a particular category in our manually labeled dataset of 2,500 comments. Finally, the \emph{Avg. Rating} for the category was computed by taking the average of \emph{User perception} and \emph{Sampled user rating} for that category.  The rows in Table~\ref{table:usefulness-ranking} are sorted based on this score.

Our results suggest that finding functional defects, validation, and logical issues are top priorities for our respondents. These results support findings from prior studies~\cite{bacchelli2013expectations,bosu2016process}.  By comparing this ranking against the classification scheme presented in Table~\ref{table:rubric}, we find most of the categories belonging to the `Functional' group ranking at the top. However, only 19\% of CR comments from our dataset belonged to these categories. This ratio is similar to the ones reported in prior studies~\cite{bosu2015characteristics,hasan2021using,beller2014modern}. Although `Documentation' and `Organization of code' rank among the bottom half (i.e., 9th and 10th respectively), more than 40\% of CR comments belonged to those categories. These results also support the prior observation that most CR comments are related to trivial issues~\cite{czerwonka2015code}.

We also noticed a few discrepancies between the ranking based on avg. \emph{User perception} and the one based on avg. \emph{Sampled user rating}. While avg. \emph{User perception} ranking for the `Validation' category places it eighth in the list, and below the `Documentation' category, it ranks top according to avg. \emph{Sampled user rating}. On the contrary, our respondents rated logical categories as the top, but the samples' ratings place that category at tenth.  The high standard deviation for the sample ratings suggests that although respondents consider CR identifying logical mistakes as highly useful,  some of the respondents rated the logical samples from our survey as less severe issues. 
The `Praise' and `Visual representation' categories ranked among the lowest. We also noticed the highest standard deviations (SD) for these two categories. These results suggest that developers have contradictory opinions regarding these categories, as some developers perceive those as useful, while others perceive those as less useful.
Other categories where our respondents' opinions have diverging opinions ( i.e., high SD) besides these two are  `Resource synchronization',  `Design discussion', and `Documentation'.

 \begin{boxedtext}
\textbf{Key takeaways:} \emph{While OpenDev developers consider `Functional' CR comments as most useful, less than 20\% CR comments from our sample belonged to that group. OpenDev developers also had widely varied views regarding the usefulness of comments belonging to praise, documentation, design discussion, resource synchronization, and visual representation. }
\end{boxedtext}

\section{RQ2. Which contextual and participant characteristics are associated with CR usefulness?}
\label{sec:results-rq2}

\begin{table*}
	\centering
    \caption{Contextual and participant factors associated with code review usefulness}     
	\label{table:usefulness-factors}
	\resizebox{\linewidth}{!}{
\begin{tabular}{p{2.9cm} p{5.5cm} p{7.5cm} }
\hline
\textbf{Attributes}            & \textbf{Description} & \textbf{Rationale} \\ \hline
\multicolumn{3}{l}{\textbf{Participant attributes}}\\ \hline
File commit count (COMC)*   & The number of times a person has made changes to a file. & With multiple prior changes, a person's awareness of the file grows and therefore odds of useful reviews~\cite{bosu2015characteristics,rahman2017predicting})\\
Weighted recent commit (WRC)     & If a file has a total of n prior commits and the author has made three of the prior n commits (e.g., i,j,k), then: $WRC = \frac{(i+j+k)}{1+2+...+n} = \frac{2(i+j+k)}{n(n+1)}$ & Recent committers of a file may have a better understanding of its current design~\cite{thongtanunam2017review}. \\
Mutual reviews (MR)            & Number of reviews the particular author and reviewer have performed mutually. & High MR indicates more mutual understanding between the author and reviewer, so the reviewer may understand the strengths and weaknesses of the author and may provide a useful review~\cite{bosu2016process}. \\
Prior file review count (PFRC)*   & The number of times a person has reviewed the current file. & With multiple prior reviews, a person's awareness of the file grows and therefore odds of useful reviews~\cite{bosu2015characteristics,rahman2017predicting,hasan2021using})\\
Reviewer code share (RCS)       & The ratio of the number of code lines written by the reviewer and the total number of lines in the code. & If the reviewer has a higher contribution in the file, then he may have a better understanding of the context and can provide a useful review~\cite{bosu2015characteristics,rahman2017predicting,hasan2021using}.\\
Review experience (PRC)        & Number of reviews the current reviewer has performed. & Reviewer review experience may have an association with review quality~\cite{hasan2021using}.\\
Reviewer commit share (RCSH)     & Ratio of the number of commits made by the reviewer and the total number of commits for the current file. & If a reviewer makes a large number of commits, then she has a higher codebase understanding and provides useful reviews~\cite{thongtanunam2017review}. \\
Coding experience (RCE) & Number of files the reviewer has submitted in the codebase as an author. & If the reviewer has higher project experience,  he may have a better understanding of the entire code base\cite{kononenko2015investigating,hasan2021using}.  \\ 
Reviewer project tenure (RPT)* & Tenure of the reviewer in number of months. & Newcomers of a project are primarily added for knowledge dissemination and are less likely to write useful reviews~\cite{bosu2015characteristics}.
\\ \hline

\multicolumn{3}{l}{\textbf{Contextual attributes}}\\ \hline
Patchset number (PN)           &      The number of total patches of a review.      &     Higher PN indicates a review going through more iterations due to useful feedback \cite{bosu2015characteristics,hasan2021using}.     \\
Comment volume (CV)            & Ratio between the number of comment lines and the total number of lines in the current file. & If a file has a high CV, then the code may become easier to understand, and the possibility of providing useful reviews increases. \\
Review interval (RI)           & The interval between the submission of the file to the review tool and the final review outcome & Time required to evaluate a file has an association with the product quality \cite{thongtanunam2017review,thongtanunam2015investigating,hasan2021using}.\\
Code churn (CCR)                & The total number of lines that have changed (added and deleted lines). & If the number of changed lines is higher, there is a higher probability of having defects \cite{nagappan2005use,nagappan2007using}. Identification of defects is considered useful by authors. \\
Is bug fix? (IBF)                & Whether the current patch is for fixing bugs or not. & Bug fixes are often associated with defect-prone files~\cite{thongtanunam2017review}. Reviewers may have higher opportunities to identify more bugs during reviews for bug fixes. \\
Is new file? (INF)               & Whether the current file is added to the patch for the first time or not. & A new file may contain more issues than a file that went through a review process previously. \\
Directory under review (DUR)    & How many directories have been affected for making current modifications &     Higher number of directories are indicators of tangled code changes~\cite{dias2015untangling}, which are difficult to review.      \\
Cyclomatic complexity (CCY)     & The cyclomatic complexity described by McCabe \cite{mccabe1976complexity} & Code comprehension difficulty increases with complexity, and therefore reduces the likelihood of useful reviews.\\
File under review (FUR)*         & How many files has been affected for making current modification &  Larger patches are time-consuming to review. As a result such patches often get cursory reviews.         \\
Current loc (CL)               & Number of lines of code currently available in the file &  Due to additional required efforts for comprehension, larger files may receive cursory reviews. \\ 
Patch description length (PDL)               & Number of words to describe the patch &  Patch description helps reviewers understand the objective of the change, and may help avoid unnecessary questions. \\ 
Readability of patch description (PDR)               & Flesch reading ease~\cite{kincaid1975derivation} score of patch description &  Patch description should be easy to understand to assist reviewers' comprehension \\ \hline
\multicolumn{3}{l}{* - attributes also investigated in Bosu et al.'s study~\cite{bosu2015characteristics}.}
\end{tabular}
}

\end{table*}

To answer this question, we selected a total of  21 attributes based on prior code review studies \cite{bosu2015characteristics,rahman2017predicting,hasan2021using,mcintosh2016empirical,fukushima2014empirical}. A total of 9 attributes are related to review participants and the remaining 12 are related to the review context. 
For each attribute, Table~\ref{table:usefulness-factors} provides a short definition and a brief rationale on why this attribute may influence the usefulness of CR comments. In a similar investigation,  Bosu \textit{et} al. selected four participant factors and two contextual factors. We selected three of the four participant factors from their study, as  OSS projects do not have the `same team' concept as Microsoft.  We selected one of the two contextual factors from Bosu \textit{et} al. We do not consider file extension in our analysis, as more than 90\%  of the sample belongs to one extension (i.e., `py'). 

We would also like to mention that we do not include any attributes related to the CR comment text due to our design constraints. Based on the results of RQ1.A, linguistics attributes of a review seem good attributes to be included in our analyses. However, we created our labeled dataset before the survey since we needed samples for the survey design. Our manual labeling rubric did not consider linguistics aspects. Judging non-technical aspects such as `respectfulness,' `understandable,' and `constructive criticism' depends on the participants; therefore, a third-party rater would not have judged those accurately.

We use Python scripts and MySQL queries to compute those attributes for the 2,500 manually labeled CR comments from the OpenDev Nova project. 
After removing the redundant variables,  we use this dataset to develop two regression models, the first one is a  Linear Regression Model (LR) to investigate how these factors the degree of usefulness achieved in a CR (RQ2.A), and the second one is a Multinomial Logistic Regression (MLR) model to identify how these factors associate with identification of functional defects.
The following subsections detail our analysis approach and results for the two sub-questions for RQ2.

\subsection{RQ2.A: How are various contextual and participant characteristics associated with the  degree of usefulness achieved in a code review?}

To answer this question, we trained a linear regression (LR) model using the \code{glm} function from the \code{stats} library in R. 
In an LR model, if the dependent variable is $Y$ and the independents are $X_1,X_2,X_3,...,X_n$, then  \code{glm} tries to fit a curve of the form $Y = \beta_0 + \beta_1X_1 + \beta_2X_2 + \beta_3X_3 + ... + \beta_nX_n$ on the dataset. Here $\beta_0$ is the intercept and $\beta_i$ is the coefficient of the independent variable $X_i$. The $\beta_i$, also indicates the change in the dependent's value if $X_i$ changes by one unit, while all the other independents remain constant. Therefore, a positive $\beta_i$ indicates a positive association of $X_i$ with $Y$ and vice versa.  The following subsections detail our model training, evaluation, and results.

\subsubsection{Model training and evaluation}
\label{sec:data-preparation}
We noticed highly skewed distributions for eight out of the 21 selected attributes. The list of skewed variables includes COMC, RI, CCR, RCE, MR, PRC, PFRC, and CL. Therefore, we applied log transformations (i.e., \code{$log_{10}$}) to model those variables' associations more accurately with our dependent variables than without such transformations \revision{\cite{osborne2002four,changyong2014log}}. 

Multicollinearity problems occur when several variables in a multivariate regression have a high degree of association with one another. The impacts of individual variables may not be accurately detected by a model trained with multicollinear factors~\cite{mansfield1982detecting}.
We used Sarle's Variable Clustering (VURCLUS) method to find highly correlated  factors~\cite{sarle1990sas}. We constructed a hierarchical cluster representation of the independents using Spearman's rank-order correlation test. Based on Hinkle \textit{et} al.'s  suggestion~\cite{hinkle1998applied} we chose the correlation coefficient $|rho| = 0.7$ as the cutoff. Only one explanatory variable from a cluster that contained multiple explanatory variables with $|rho| \geq 0.7$ was selected. 
After evaluating correlation and redundancy, we chose the most effective set of factors for our model using Jiarpakdee et al.'s~\cite{jiarpakdee2018autospearman} \code{Autospearman} R library.

\begin{figure}
\centering  \includegraphics[width=\linewidth,trim={0 2.6cm 0 2cm},clip]{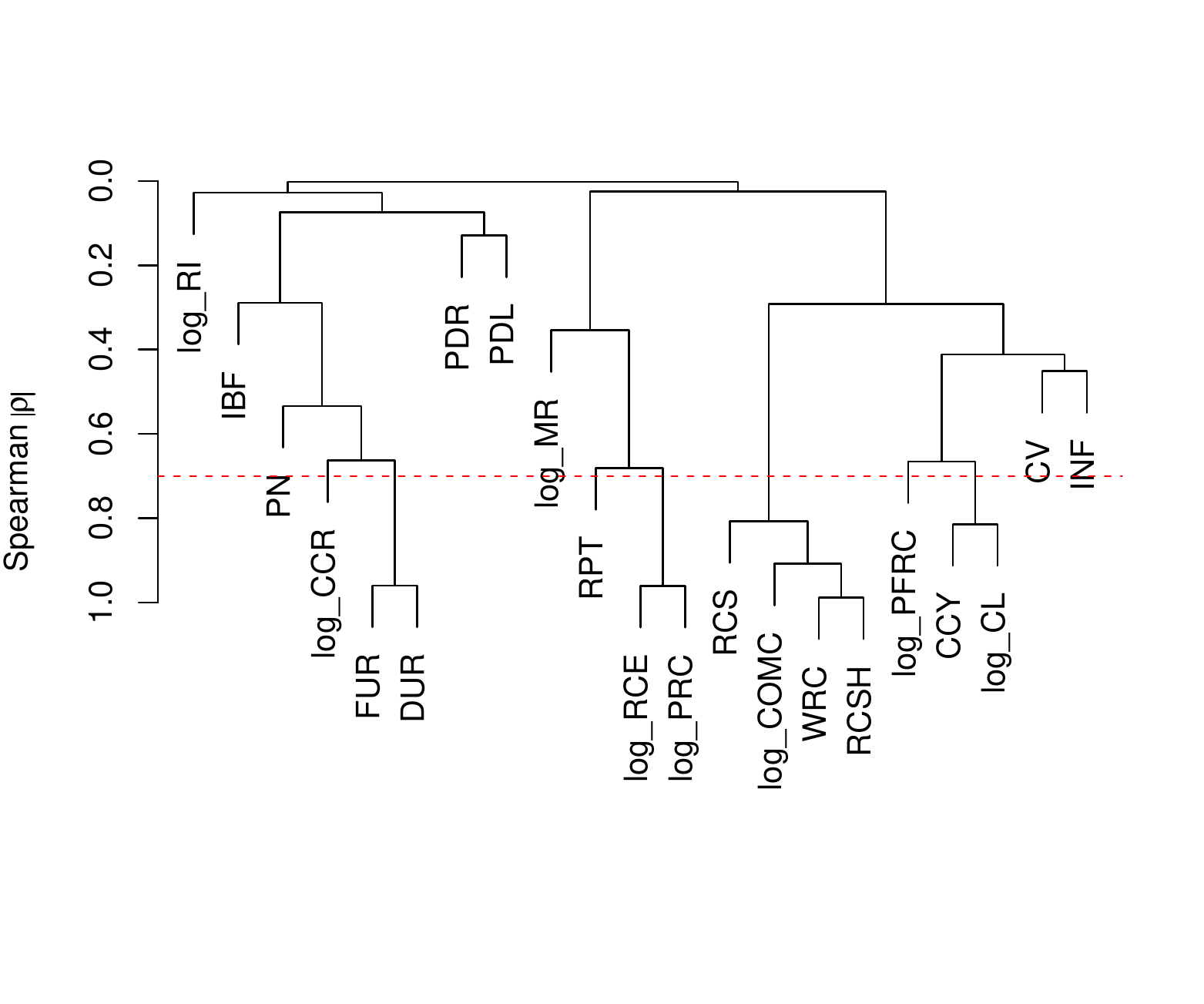}
\caption{Hierarchical structure among the independent variables according to Pearson correlation}
\label{fig:hierarchical_clustering}
\end{figure}

Figure ~\ref{fig:hierarchical_clustering} shows the hierarchical structure of the independent variables based on correlation analysis. We identified four highly correlated clusters. 
From the (RCS, WRC,  RCSH, log\_COMC)  cluster, \code{AutoSpearman} keeps RCS. From the (log\_CL,  CCY) cluster CCY survives.
 From (log\_RCE, log\_PRC) group, log\_RCE survives.       
 Finally, from the  (FUR, DUR), FUR survives. 
 After excluding these six variables, we use the remaining 15 variables to construct our regression models.  In our analysis, the dependent variable is a continuous variable named \emph{usefulness\_score} and the independents are  15 of the surviving factors from Table \ref{table:usefulness-factors}.  

For each CR comment from our labeled dataset \revision{(there are 2,500 CR comments in the labeled dataset)}, we compute its \emph{usefulness\_score} according to the following formula.
\begin{enumerate} [label=\roman*.]
    \item If a comment is labeled as `Not useful', we assign a `0' \emph{usefulness\_score}.
    \item If a comment is labeled as `Useful', we check its assigned category. From the \emph{Sampled user rating} column of Table~\ref{table:usefulness-ranking}, we take the score obtained for this category and that score is its \emph{usefulness\_score}.
\end{enumerate}

While we had two other possible options, i.e., `User perception' and `Avg Rating',  we found potential issues with those choices for usefulness scores.  We could not validate that our respondents indeed fully comprehend a CR comment category based on the brief definitions included in our survey. As `Sampled user ratings' are obtained based on actual ratings of reviews, those are more accurate than the other two. 

We assess the goodness of fit for the trained model using McFadden’s $R^2$, which is 0.026 ($p<0.001$). 
We performed a log-likelihood ratio to compare our fitted model with a null model. Our test results suggest ( $\chi^2$ = 69.9,   $p-value<0.0001$ ) significant explanatory power of the model over a null model. 
We acknowledge that the  $R^2$ value achieved by this model is lower than what we had anticipated. This lower value may be due to two primary reasons. First,  in a randomly selected code review sample such as ours, the majority of comments belong to trivial issues such as `Documentation' and `Naming convention'. As these comments are more likely to be participant or context agnostic, a model trained using randomly chosen samples is less likely to achieve high explanatory power. Second, approximately 77\% CR comments in our dataset are useful ones (i.e., scores higher than zero). As a result, we notice a high intercept value (i.e.,  3.02). 
Instead of a randomly sampled one, if we had a dataset with higher ratios of `Not useful' comments (i.e., score =0), a model trained with such a dataset would have achieved a higher $R^2$. Regardless, since our goal is to develop an inferential regression model, not a predictive one, a low  $R^2$ score does not invalidate the obtained insights. Moreover, our model significantly outperforms a null model. 

\begin{table*}
	\centering
    \caption{Results of or linear regression model to identify how contextual and participant factors associate with the degree of CR usefulness. $\beta$ in \bluebg{blue} background indicates a significant positive association, and \pinkbg{red} indicates a significant negative association.}    
	\label{table:glm_parameters}
	\begin{tabular}{l|r|r|r|l}
\hline
\multicolumn{1}{c|}{Attribute} & \multicolumn{1}{c|}{ Co-efficient ($\beta_i$)} & 95 \%Confidence interval for $\beta_i$  & $\chi^2$  & \multicolumn{1}{c}{p-value} \\ \hline
log\_MR  &   \negative{-4.02e-02}  &  [ -7.64e-02  ,  -4.03e-03 ] & 4.745 &  \textbf{0.0295} $^{*}$ \\ \hline 
 RPT  &   \negative{-5.92e-03}  &  [ -1.08e-02  ,  -1.01e-03 ] & 5.590 &  \textbf{0.0181} $^{*}$ \\ \hline 
 log\_PFRC  &   1.69e-02  &  [ -5.18e-02  ,  8.56e-02 ] & 0.232 &  0.6301  \\ \hline 
 RCS  &   -6.37e-03  &  [ -2.72e-02  ,  1.44e-02 ] & 0.359 &  0.5488  \\ \hline 
 log\_RCE  &   \positive{1.34e-01}  &  [ 6.90e-02  ,  1.99e-01 ] & 16.287 &  \textbf{0.0001} $^{***}$ \\ \hline 
 PN  &   1.67e-03  &  [ -7.19e-03  ,  1.05e-02 ] & 0.136 &  0.7122  \\ \hline 
 CV  &   \negative{-3.66e-01}  &  [ -5.69e-01  ,  -1.63e-01 ] & 12.471 &  \textbf{0.0004} $^{***}$ \\ \hline 
 FUR  &   -1.30e-03  &  [ -8.39e-03  ,  5.79e-03 ] & 0.129 &  0.7186  \\ \hline 
 log\_RI  &   \negative{-4.83e-02}  &  [ -7.80e-02  ,  -1.86e-02 ] & 10.154 &  \textbf{0.0015} $^{**}$ \\ \hline 
 log\_CCR  &   2.99e-02  &  [ -3.52e-02  ,  9.51e-02 ] & 0.812 &  0.3677  \\ \hline 
 IBF  &   1.88e-03  &  [ -1.55e-01  ,  1.59e-01 ] & 0.001 &  0.9813  \\ \hline 
 INF  &   1.53e-01  &  [ -1.09e-01  ,  4.15e-01 ] & 1.315 &  0.2517  \\ \hline 
 CCY  &   4.20e-05  &  [ -2.08e-04  ,  2.92e-04 ] & 0.108 &  0.7414  \\ \hline 
 PDR  &   -3.72e-04  &  [ -3.48e-03  ,  2.74e-03 ] & 0.055 &  0.8145  \\ \hline 
 PDL  &   -5.92e-04  &  [ -1.37e-03  ,  1.85e-04 ] & 2.229 &  0.1356  \\ \hline  
\multicolumn{4}{p{10cm}}{*** , **, and *  represent statistical significance at $p <$ 0.001, $p <$ 0.01, and $p <$ 0.05 respectively.} \\

\end{tabular}    
\end{table*}

\subsubsection{Model analysis}
Table \ref{table:glm_parameters} reports the co-efficient ($\beta_i$) for the surviving variables, 95\% confidence interval estimates for the coefficients,  and statistical significance of the associations (i.e. p-value).
A negative coefficient value indicates if this variable increases, the CR usefulness score decreases and vice versa. 
Each variable's $\chi^2$ value is the difference between $-2\;Log\;Likelihood$ of the full model and a model constructed by dropping that particular variable under discussion. A higher $\chi^2$ value indicates a higher explanatory power added by an independent. 
Our model's parameter estimates suggest a total of five independent variables having statistically significant associations ($p-value<0.05$ and presented in bold format). Among those, log\_RCE ( log of reviewers' coding experience) has the highest level of influence with a positive coefficient. This result suggests that experienced coders, who have significant project knowledge due to a large number of prior code changes, are more likely to provide useful comments. 
On the contrary, RPT (reviewer's project tenure has a negative association). As this association is contradictory to RCE, we further investigated the reasons. We computed persons' tenures from their first interaction with the project. We found many casual contributors who have long tenures but have made very few commits. Less useful CR comments by these casual contributors contribute to the negative association between CR usefulness and RPT.
Contrary to our initial assumptions from Table \ref{table:usefulness-factors}, both log\_MR (mutual reviews) and log\_RI (review interval) also have significant negative associations. 
Although `Quid pro quo' reviews are common~\cite{bosu2016process}, negative $\beta$ for log\_MR discourages such practices.  Our results from RQ2.B also provide more explanation of this finding.
We also notice that delayed reviews do not indicate more useful reviews.
If a review takes longer than usual, the reviewer may be overloaded and have done a cursory review to avoid further delays.
Finally, we noticed CV (comment volume) lowering CR usefulness. Our investigation found that if a file includes higher ratios of comments, reviewers often include suggestions to improve those with better wording, explanations, and grammar fixes. Since these types of suggestions have lower usefulness scores,  CV has a negative association.

 \begin{boxedtext}
\textbf{Key takeaways:} \emph{i) While a reviewer's coding experience positively associates with code review usefulness, their number of mutual reviews with the author, code review interval, comment volume in the file, and selection of casual contributors have opposite associations.}
\end{boxedtext}

\subsection{RQ2.B: RQ2.B: How are various contextual and participant factors associated with identifying functional defects?}
When the dependent variable is nominal with more than two levels, MLR allows modeling the likelihood of a particular outcome occurring given the values for a set of independents. 
MLR is a Maximum Likelihood Estimator, where the log odds of the dependent variable are captured as a linear combination of the independents.
Since our dependent variable has five levels (i.e., category of CR comment), MLR is a right fit for our analyses. MLR also  has the following advantages:

\begin{itemize}
  \item MLR can operate in cases where variables are not normally distributed.
  \item MLR can be applied when there is no linear relationship between independent and dependent variables \cite{bayaga2010multinomial}.
  \item For MLR analysis, independent variables can be both discrete and interval (continuous) type, whereas most other analysis requires independent variables to be continuous \cite{bayaga2010multinomial}.
  \item MLR does not require inclusion of error terms that are normally distributed \cite{bayaga2010multinomial}.
\end{itemize}

Moreover, MLR allows modeling the likelihood of changing to a particular outcome from the existing outcome if a particular independent variable changes.  Therefore, with MLR, we can set a reference CR comment category and analyze the likelihood of another  CR comment category if particular independent changes. Following subsections detail our model construction and analysis approach.

\subsubsection{Model training and evaluation}
For our MLR model, we set  \emph{comment\_group} as the dependent variable, which denotes a CR comment's group listed in Table~\ref{table:rubric}. As our categorization scheme listed in Table~\ref{table:rubric} divides CR comments into five categories, this categorical variable has five possible values. We set the `Functional' category as the reference since comments belonging to this category are a top priority. \revision{We excluded the 106 comments belonging to `Others' for this analysis.}
 We use the \code{multinom} function of the \code{nnet} R package to construct our MLR model. We estimate our models' goodness of fit using Nagelkerke $Pseudo\;R^2$~\cite{nagelkerke1991note}, which is $0.221$ for our model.
Our Log Likelihood test's results ($\chi^2$ =567.67, $p<0.001$) suggest that our model is significantly better than a Null model. Our model allocates a total of 60 degrees of freedom for the final model, which is significantly lower than the maximum $\frac{2500}{15} = 166$ recommended degrees of freedom by Harrell Jr.~\cite{harrell1984regression}. Therefore, our model is not overfitted.

\begin{table*}
	\centering
    \caption{Identify the factors that distinguish \emph{Functional} class from other classes. $OR$ in \bluebg{blue} background indicates a significant higher likelihood of  \emph{Functional} comments if that variable increases by one unit and \pinkbg{red} indicates a significant lower likelihood  for \emph{Functional} ones.} 
	\label{table:parameter_estimate}
	\begin{tabular}{l|r|l|r|l|r|l|r|l}
\hline
\multicolumn{1}{c|}{\multirow{2}{*}{\textbf{Attribute}}} & \multicolumn{2}{c|}{\textbf{Discussion}} & \multicolumn{2}{c|}{\textbf{Documentation}} & \multicolumn{2}{c|}{\textbf{False Positive}} & \multicolumn{2}{c}{\textbf{Refactoring}} \\ \cline{2-9} 
\multicolumn{1}{c|}{}                                        & \multicolumn{1}{c|}{\textbf{OR}}                                  & \multicolumn{1}{c|}{\textbf{$p$}}                                  & \multicolumn{1}{c|}{\textbf{OR}}                                 & \multicolumn{1}{c|}{\textbf{$p$}}                                 & \multicolumn{1}{c|}{\textbf{OR}}                                 & \multicolumn{1}{c|}{\textbf{$p$}}    & \multicolumn{1}{c|}{\textbf{OR}}                                 & \multicolumn{1}{c}{\textbf{$p$}}      \\ \hline

log\_MR & \negative{1.11} & \textbf{0.028}$^{*}$ & \negative{1.09} & \textbf{0.047}$^{*}$ & \negative{1.13} & \textbf{0.009}$^{**}$ & 0.96 & 0.289\\ \hline 

RPT & 1.00 & 0.561 & 1.00 & 0.475 & 1.01 & 0.296 & 0.99 & 0.310\\ \hline 

log\_PFRC & 1.09 & 0.281 & 1.05 & 0.545 & 0.95 & 0.471 & 1.03 & 0.645\\ \hline 

RCS & \positive{0.45} & \textbf{0.007}$^{**}$ & 0.94 & 0.537 & 0.82 & 0.743 & 0.96 & 0.628\\ \hline 

log\_RCE & 0.86 & 0.068 & 1.08 & 0.332 & \positive{0.84} & \textbf{0.022}$^{*}$ & 1.02 & 0.772\\ \hline 

PN & 1.00 & 0.831 & 1.00 & 0.704 & 1.00 & 0.945 & \negative{1.02} & \textbf{0.042}$^{*}$\\ \hline 

CV & 1.70 & 0.079 & \negative{10.25} & 0.00$^{***}$ & \negative{4.21} & \textbf{0.00}$^{***}$ & 0.89 & 0.705\\ \hline 

FUR & \positive{0.98} & 0.016$^{*}$ & \positive{0.97} & \textbf{0.000}$^{***}$ & \positive{0.98} & \textbf{0.027}$^{*}$ & \positive{0.97} & \textbf{0.001}$^{**}$\\ \hline 

log\_RI & 1.01 & 0.677 & 0.94 & 0.057 & 1.06 & 0.084 & 1.00 & 0.887\\ \hline 

log\_CCR & \negative{1.27} & \textbf{0.002}$^{**}$ & \negative{1.29} & \textbf{0.000}$^{***}$ & 1.08 & 0.317 & \negative{1.23} & \textbf{0.005}$^{**}$\\ \hline 

IBF & 1.04 & 0.846 & 0.97 & 0.878 & 0.97 & 0.871 & 1.14 & 0.483\\ \hline 

INF & 1.20 & 0.645 & 1.31 & 0.455 & 0.93 & 0.844 & 0.90 & 0.795\\ \hline 

CCY & 1.00 & 0.05 & 1.00 & 0.0618 & 1.00 & 0.137 & 1.00 & 0.722\\ \hline 

PDR & 1.00 & 0.483 & \positive{0.99} & \textbf{0.026}$^{*}$ & 0.99 & 0.215 & 1.00 & 0.552\\ \hline 

PDL & 1.00 & 0.855 & 1.00 & 0.137 & 1.00 & 0.171 & 1.00 & 0.537\\ \hline

\multicolumn{9}{p{11.5cm}}{*** , **, and *  represent statistical significance at $p <$ 0.001, $p <$ 0.01, and $p <$ 0.05 respectively.}

\end{tabular}
\end{table*}

\subsubsection{Model analysis}

Table~\ref{table:parameter_estimate} shows the results of our MLR models with \emph{Functional} comment group as the reference. The $OR$ column under a particular category indicates the odds of that category occurring instead of the `Functional' if a variable listed under the first column changes by one unit.  $OR > 1$ indicates a higher probability of the target class than the reference (i.e., `Functional') if the associated attribute increases and vice versa.
For example, the $OR$ value for the variable log\_MR \revision{(i.e., log of mutual code reviews)} under `Discussion' is 1.11 with a $p$ value of 0.0285 (i.e., significant at $p<0.05$). Therefore, if log\_MR increases by one unit, the likelihood of a `Discussion' instead of `Functional' increases by 1.11. Similarly, log\_MR is also associated with a higher likelihood of both `Documentation' and `False positive.' These results shed further light on our findings from RQ2.A, which found a negative association between log\_MR and CR usefulness score.

On the other hand, under the `Discussion,' $OR$ =0.98 for the variable FUR \revision{(file under review)}  indicates that if the number of files under review increases, the likelihood of comments identifying `Functional' defects also increases. Similar associations are seen between FUR and the other three categories as well. These findings may not be surprising since prior studies have found changes involving a higher number of files are more likely to include defects~\cite{Bosu-et-al-FSE:2014}. Hence, such changes are more likely to receive `Functional' comments.

A significant odds reduction for `Discussion' with RCS (reviewer's codeshare) suggests that the likelihood of discussion over `Functional' decreases if the reviewer's ownership of the file increases.
Our results suggest that  log\_CCR (i.e., log of code churn) decreases the likelihood of `Functional' over all four categories. These results support prior findings that reviewers opt for shallow reviews for larger code changes~\cite{bacchelli2013expectations}.
We notice a negative association between log\_RCE (i.e., log of a reviewer's commit experience) and `False positive,' which suggests that the likelihood of a `False positive' decreases with increased commit count by the reviewer.

A higher likelihood of `Refactoring' comments with increased `PN' \revision{(patchset number)} suggests that if the number of iterations in a CR increase, the later patches are more likely to receive `Refactoring' suggestions than `Functional' ones.
We notice the highest change in OR value (i.e., 10.25) for CV \revision{(comment volume)} under `Documentation.' This result suggests that if comment volume in a file increases, the odds of suggestions to improve those documentation increases drastically. Moreover, a CV increment increases the likelihood of false positives as well. A lower effort requirement during reviews for suggesting documentation improvements than for `Functional' ones may be a possible cause.

Surprisingly, we notice PDR (the patch description's readability) improves the odds of `Functional' over `Documentation.' These results suggest that well-described patches are more likely to receive `Functional' than `Documentation' comments. 
Finally, although we noticed log\_RI \revision{(i.e., log of review interval)} negatively associating with CR usefulness score in RQ2.A, we did not find any significant association here. These results indicate that delayed reviews do not increase the odds of any particular comment category. Those are more likely due to overloaded reviewers providing delayed yet shallow reviews.

 \begin{boxedtext}
\textbf{Key takeaways:} \emph{The number of mutual reviews between the author and a reviewer,  the total number of lines added /modified in a change, and the ratio of lines that are comments are negatively associated with receiving comments identifying `Functional' defects. 
The odds of functional defects and their identification increase with the number of files under review. 
Experienced committers of the projects are less likely to author invalid suggestions.
Delayed reviews do not increase the odds of any particular category of comment.  }
\end{boxedtext}

\section{Discussion and Implications}
\label{sec:implications}
\textbf{1. Comparison with Bosu et al.\cite{bosu2015characteristics}: } 
Similar to Bosu \textit{et} al. ~\cite{bosu2015characteristics}, we found `Functional' defect identification as the top priority for the CR participants. \revision{As the study of Bosu et al. was conducted in a commercial setup (Microsoft), by contrasting the findings of the two studies, we can get a picture of the perspective differences for commercial and OSS projects.} Our results deviate from \revision{the study of Bosu et al.} among several key aspects as described in the following.

\begin{itemize}
    \item Participants from  Bosu \textit{et} al. did not consider questions to understand the implementation as `Useful'.  However, most of our respondents consider questions as useful, with its average `sample user rating' ranking fourth and `Avg. Rating' ranking sixth.
    
    \item Similarly, participants from Bosu \textit{et} al.'s study consider `Praise' and `Design discussion' as `Not Useful'. However, the majority of our respondents consider `Design Discussion' as useful.     We found `Praise' as one of the categories with the highest standard deviation, which indicates developers' opinions vary widely regarding this category. 
    
    \item Bosu \textit{et} al. found that a reviewer's prior experience with the file under review, either as a reviewer or as an author, positively associates with the likelihood of providing useful reviews. However, we found none of the factors to measure these experiences (i.e., RCS, log\_PFRC, and log\_COMC  ) having significant associations with usefulness scores in our study.
    
    \item Bosu \textit{et} al. found CR usefulness drops as the number of files under review increases. However, we did not notice any such association. On the contrary, we found the likelihood of `Functional' defects increasing with the number of files. 
    
    \item Bosu \textit{et} al. found a positive association between the likelihood of useful reviews and reviewers' Microsoft tenure. However, we found a negative association between project tenure and useful reviews. We found the presence of casual contributors~\cite{pinto2016more}, which does not apply to Microsoft, as the reason behind this difference.  
\end{itemize}

\vspace{4pt}
\noindent \textbf{2. Comparison with  Kononeko \textit{et} al.~\cite{kononenko2015investigating,Kononenko-2016}}

\revision{Kononenko \textit{et} al. conducted a study on Mozilla to understand whether people and participation have an influence on review quality as measured by missed defects~ \cite{kononenko2015investigating}. 
Results of their study suggest that the number of files in a patchset is positively associated with post-review defects, while the reviewer's experience with the project has the opposite association. While we have a different measure, similar to Kononeko \textit{et} al., we found the number of files negatively associated with  review usefulness. However, we did not notice any significant association between the experience measures and review usefulness.
In a subsequent study, Kononenko \textit{et} al. conducted a survey of 88 Mozilla core developers to understand their perspective on the code review quality, factors influencing their code evaluations, and challenges encountered during reviews\cite{Kononenko-2016}. One of their research questions, where they investigated the characteristics of a well-done code review, resembles our research question RQ1.A. Although several perspectives of the Mozilla core developers, such as finding defects, being constructive, being clear and thorough,  reviewing on time, and  sharing knowledge are similar to ones expressed by the OpenDev developers, we also found several new aspects of review usefulness, which include being respectful, appreciating good work, facilitating better designs, and improving project maintainability. }

\vspace{4pt}
\noindent \textbf{3. Comparison with other studies investigating CR usefulness:}
Hasan \textit{et} el. partially replicated Bosu \textit{et} al.'s study at another industrial context (i.e., SRBD). The results of their study concur with ours that `Functional' defects are top priorities and developers have mixed opinions regarding the usefulness of `Praise', `Questions', and `Documentation'. 
Rahman \textit{et} al. investigated the relationship between developer experience and CR usefulness. Our results concur with their findings that code-commit experience has positive associations. However, contrasting their findings we do not find any significant association between CR usefulness and authorship/reviewership for the file under review.

\vspace{4pt}
\noindent \textbf{3. Reviewer selection: }
The results of our study suggest that CR usefulness decreases if the number of mutual reviews increases. Our results further suggest that such pairs are likelier to engage in discussions, identify false issues, or suggest trivial documentation-related changes. With a higher number of mutual reviews, two persons are more likely to be aware of each others' strengths and weaknesses and therefore overlook the author's areas of strength. Moreover, such reviews become more predictable and do not bring any fresh perspectives. Although an author is more likely to receive acceptance votes from such peers~\cite{thongtanunam2020review}, the results of this study discourage `Quid pro-Quo' reviews and recommend rotating reviewers, if possible. Moreover, such rotations will also help knowledge decentralization. 

The results of our study also suggest that the likelihood of useful reviews increases with the reviewer's commit count. Therefore, experienced authors of a project are the best reviewers, who not only provide useful feedback but also are significantly less likely to write invalid ones. However, this approach has drawbacks due to knowledge centralization as well as overloading experienced contributors. Therefore, we recommend adding such contributors as reviewers for critical code changes. 

We also found reviewers who have long tenure with the project but have not contributed many commits (i.e., `Casual contributors'~\cite{pinto2016more}) being associated with low CR usefulness.  While authors may invite such contributors as reviewers for various reasons, they should also include qualified reviewers in such cases to ensure adequate scrutiny.

\vspace{4pt}
\noindent \textbf{4. Recommendation to reviewers: } CR comments belonging to the `Functional' group are most useful to authors. Therefore, a reviewer should focus the most on providing such comments. Comments belonging to the `Refactoring' group rank second. However, `Documentation', nit-picking, and style issues that can be identified using static analysis tools should get the lowest priority. Finally, instead of keeping an author waiting and providing a shallow review, a reviewer should avoid being in such a scenario by rejecting review invitations if they cannot review on time or spend adequate effort.

\vspace{4pt}
\noindent \textbf{5. Recommendation to authors:} 
Our results also suggest that the readability of patch description improves the likelihood of `Functional' comments than `Documentation' ones. Therefore, an author should create a meaningful and readable description of the patchset during review preparation. 
On the contrary, CR usefulness decreases if comment volume increases. Although most authors do not find suggestions to improve `Documentation' useful, they are more likely to receive those if they write more comments. `Documentation' comments are low-hanging fruits, and reviewers are more likely to produce those if they do not have time for thorough reviews. Therefore, if an author finds `Documentation' comments useless, we recommend following the Agile Manifesto guideline~\cite{fowler2001agile} by writing self-documenting code and writing documentation only when necessary. 

Similar to prior studies~\cite{bacchelli2013expectations}, we found changes with higher code churns are less likely to receive CR comments identifying `Functional' changes. Therefore, we recommend committing smaller incremental changes.

Finally, delayed reviews not only frustrate authors but also such reviews are less likely to be helpful. Our results indicate that delayed reviews are less likely to be the results of thorough reviews but more likely due to overloaded reviewers or other factors. Therefore, an author should invite an alternative instead of waiting for a reviewer who is already late beyond reasonable expectation.

\vspace{4pt}
\noindent\textbf{6. Using appropriate language: } CRs are most useful to developers when participants discuss code issues constructively and empathetically. Constructive conversation helps to maintain a healthy relationship among the teammates \cite{sarker2020benchmark}. Approximately one-third of respondents indicated appropriateness of the language as an important factor in judging CR usefulness. Inappropriate language may shift the attention from the actual issue (i.e., code under review) to the participant's personal characteristics and, therefore, not only degrade CR's usefulness but can instigate conflicts. Prior studies also have found that inappropriate languages hinder diversity, equity, and inclusion initiatives by disproportionately hurting the participation of women and other minorities~\cite{gunawardena2022destructive}. 

\vspace{4pt}
\noindent\textbf{7. Recommendation to researchers: }
The deviations between our results and the ones reported in prior studies~\cite{bosu2015characteristics,rahman2017predicting} suggest that more replications are crucial to understanding how various project or organizational factors influence CR usefulness. Although authorship / reviewership experience for the artifacts under review is the most favored attribute for reviewer recommendation systems~\cite{rahman-correct,zanjani2015automatically,thongtanunam2015should}, our results indicate no significant association with those with CR usefulness scores. Both  Bosu \textit{et} al.  Rahman \textit{et} al. found a non-linear relationship between CR usefulness and those experience measures, where CR usefulness increases linearly with experience measures at lower values but plateaus beyond a certain threshold, such as 5-15. As we log-transformed those experience measures (i.e., PFRC and COMC), we expected linear relationships between usefulness score and (log\_COMC, log\_PFRC). However, we found no significant association between file experience measures and CR usefulness. Although log\_COMC was dropped from the model due to multicollinearity with RCS, we did not notice any significant association with RCS either. Hence, if we had trained models with log\_COMC instead of RCS, we would have seen a similar association as the one observed for RCS. This result raises whether these attributes are good candidates for building reviewer recommendation systems. While it is premature to make a definite recommendation based on a single study, we recommend more investigations to identify context-specific appropriate attributes for reviewer recommendation systems.

We also notice that attributes that increase the likelihood of the `Functional' group differ from those increasing `Discussion' or `Documentation.' While existing multi-objective reviewer recommendation systems~\cite{mirsaeedi2020mitigating,asthana2019whodo} consider expertise, workload, and knowledge distributions, identifying different categories of issues has not been explored. We recommend considering this aspect as another objective for multi-objective reviewer recommendation systems.

\section{Related Work}
\label{sec-related-works}
Besides finding defects, known benefits of CR include: improving project maintainability, maintaining code integrity, improving relationships between the participants~\cite{bosu2016process}, preventing security defects~\cite{Bosu-et-al-FSE:2014,paul2021security} and spreading knowledge, expertise, and development techniques among the review participants~\cite{bacchelli2013expectations,rigby2013convergent}. Hence, CR  has achieved\\ widespread adoption among both commercial OSS development organizations~\cite{bacchelli2013expectations,rigby2013convergent,sadowski2018modern}. Many projects mandate each code change be approved through  CR  before it can be considered for integration into the project's main code base~\cite{rigby2013convergent}. 

Despite significantly associated efforts, most of the CRs do not find bugs~\cite{czerwonka2015code}, as the majority of the CR comments involve maintainability issues that do not impact code output~\cite{bacchelli2013expectations}. Defect identification during CRs requires time commitment as well as a deeper understanding of the code context. Reviewers inadequately fulfilling these two criteria often focus more on refactoring or nit-picking issues than defect identification~\cite{bacchelli2013expectations}. 
Several studies have investigated the types of issues identified during CRs~\cite{beller2014modern,bosu2015characteristics,hasan2021using,bacchelli2013expectations,mantyla2008types} and found similar ratios across various commercial as well as OSS projects, with functional defects representing less than 20\% of the CR comments.

On a goal to improve CR effectiveness, prior studies have focused on understanding what makes a CR useful to the participants~\cite{bosu2015characteristics,hasan2021using,rahman2017predicting}. 
In this direction, Bosu \textit{et al.}\cite{bosu2015characteristics} conducted a three-stage empirical study in Microsoft.
In the first stage, they interviewed developers to understand what makes a CR useful. Based on the insights obtained from the interviews, they manually labeled a dataset CRs to train an automated model to predict useful reviews. In the third stage, they used the automated model to predict the usefulness of 1.5 million CR comments and conduct an empirical study to understand how CR usefulness varies with various factors~\cite{bosu2015characteristics}. 
The results of their empirical study suggest that the reviewer’s prior experience in changing or reviewing the artifact and the reviewer’s project experience increases the likelihood that s/he will provide useful feedback.
Hasan \textit{et} al. replicated the first two stages of Bosu \textit{et} al. study in another commercial organization and found that while developers agree on several CR usefulness criteria such as defect identification or solution approach, the same cannot be said for suggestions to improve documentation, praise, or questions asking clarification~\cite{hasan2021using}. While these two studies only focused on the notion of usefulness post-review completion, Rahman \textit{et} al. developed a classifier to predict useful CR comments using both textual and contextual features~\cite{rahman2017predicting}.

Kononenko et al. \cite{kononenko2015investigating} examined CR quality and found that personal factors such as reviewer workload, experience, and participation factors such as the number of developers involved have significant associations. 
A later empirical study at Mozilla, by the same set of authors, found that useful CRs are thorough and are influenced not only by the reviewer’s familiarity with the code but also by the perceived quality of the code itself \cite{Kononenko-2016}. Bosu \textit{et} al.'s survey of OSS and Microsoft developers suggest that several human factors such as the author's reputation, and the relationship between an author and a reviewer, also dictate CR effectiveness~\cite{bosu2016process}.
The results of Hatton \textit{et} al.' suggest that the capability to identify defects during CRs may vary widely among reviewers as the best reviewer can be up to 10 times more efficient than the worst ones~\cite{hatton2008testing}. 
Other factors decreasing CR quality include missing rationale, discussion of non-functional requirements of the solution, and lack of familiarity with existing code~\cite{ebert2021exploratory}, co-working frequency of a reviewer with the patch author ~\cite{thongtanunam2020review}, description length of a patch~\cite{thongtanunam2017review}, and the level of agreement among the reviewers~\cite{hirao2016impact}.

Several recent studies have focused on improving CR effectiveness through the automation of various CR tasks. Reviewer recommendation systems~\cite{rahman-correct,thongtanunam2015should,zanjani2015automatically,rong2022,pandya2022} focus on finding the best reviewer(s) for a given change. Since understanding changes during CRs are often time-consuming, researchers have proposed tools and frameworks to support changeset comprehension~\cite{barnett2015helping,dias2015untangling,tao2015partitioning,gomez2015visually,huang-salient}. Changeset size reduction is another automation direction, which aims to decide which change fragments are error-prone and need to be checked in detail to expedite a CR process~\cite{baum2016need}. 
Recent studies have focused on automating the reviews through ML-based models~\cite{tufan2021towards,tufano2022using,thongtanunam2022autotransform,hong2022commentfinder}.

\section{Limitations}
\label{sec:threats}

\textbf{Internal validity:}  Our sample selection is the primary threat to internal validity. While we surveyed developers from the entire OpenDev community, we selected CR comments from only the OpenDev Nova project. While Nova is the largest and most active project in this community, we cannot claim that it is an accurate representation of the entire community. However, we are unaware of any evidence regarding the contrary.

\vspace{2pt} \noindent \textbf{Construct Validity: } Our survey design and code snippet selection may be subject to biases. To mitigate the bias, we send the survey to \revision{two} experts in the Software Engineering field and modify the survey based on their feedback. We also told three student researchers to identify any ambiguous questions before sending the survey to the developers.

For our quantitative analyses (i.e., RQ2.A, RQ2.A, RQ2.B), we focused only on comment categories to measure CR usefulness since some of the secondary criteria, such as knowledge sharing and relationship formation, are difficult to \revision{evaluate by an independent rater within the limited context of a survey}. While we are aware of prior CR studies~\cite{mirsaeedi2020mitigating,chouchen2021whoreview} introducing a metric to measure knowledge sharing, we could not use those metrics since our unit of analyses is individual CR comment, as opposed to the entire CR used in those studies. 

Multicollinearity is a potential threat to regression models. Multicollinearity-related threats arise when two independent variables are highly correlated, and due to the interaction between two multicollinear variables, the interaction between an independent and the dependent variables is underestimated. To mitigate this threat, we followed Harrell Jr.~\cite{harrell1984regression}'s approach to identifying highly correlated variable clusters and picked only one variable from a cluster to train our models.

\vspace{2pt} \noindent \textbf{External Validity:} 
We divided our study objectives into two research questions. The first question aims to understand OSS developers' perspectives about the usefulness of CRs and comment categories that are considered most useful to the developers. Although we conducted our survey on OpenDev developers, most of the developers have experience working on other open-source and industrial projects. 
Therefore our findings might apply to other OSS and industrial projects. However, as OSS projects vary based on technology, norms, culture, and governance, external validity remains a threat. 

Our second research question aims to identify the factors that are associated with CR usefulness. Since this analysis includes only one project's data,   this result may not be generalized outside of our study subject. 
Manually labeling a large dataset using multiple raters is time-consuming, especially when we have 18 different choices for each instance. Therefore, we were unable to include multiple projects' datasets in this analysis.
Replications of this study are essential to derive context-specific customized CR-specific recommendations. To promote replications, we have made our survey questions, anonymized dataset, and analysis scripts publicly available.

\vspace{2pt} \noindent \textbf{Conclusion validity:} While constructing a regression model, overfitting appears to be the primary threat to conclusion validity. We followed Harrell Jr.'s recommendations to encounter this threat by allocating degrees of freedom less than (n/15), where n is the size of the dataset. For model training, we use \code{stats} and \code{nnet} packages, which are considered the gold standards for regression modeling. Hence, we do not anticipate any threat to library selections.
Finally, we noticed a low $R^2$ value for our linear regression model due to the nature of the code review dataset. However, as our goal is to train inferential models, this low measure does not invalidate obtained insights. Moreover, our log-likelihood tests found this model significantly better than a null model.

\section{conclusion}
\label{sec:conclusion}
We conducted a three-stage mixed-method study to investigate CR's usefulness among OSS developers. We manually categorized 2,500 CR comments, using those comments designed an online survey,  received 160 usable responses from OpenDev developers, combined insights obtained from the survey with our manually labeled dataset, and finally trained two regression models to provide a better understanding of what makes code review useful and the set of factors influencing CR usefulness. The results of our study suggest that a CR comment's usefulness is dictated not only by its technical contributions such as defect findings or quality improvement tips but also by its linguistic characteristics such as comprehensibility and politeness. 
While a reviewer's coding experience positively associates with CR usefulness, the number of mutual reviews, comment volume in a file, the total number of lines added /modified, and CR interval have the opposite associations. 
While authorship and reviewership experiences for the files under review have been the most popular attributes for reviewer recommendation systems, we do not find any significant association of those attributes with CR usefulness. As we find several of our results deviating from prior studies, we also recommend more investigations to identify context-specific attributes for reviewer recommendation models.

\section*{Data availability}
Our analysis scripts, and aggregated dataset are publicly available at: \\ \url{https://github.com/WSU-SEAL/CR-usefulness-EMSE}. Upon acceptance, we plan to post it on Zenodo with a permanent DOI.

\section*{Funding and Conflicts of interests/Competing interests.}

Work conducted for this research is partially supported by the US National Science Foundation under Grant No. 1850475. Any opinions, findings, and conclusions or recommendations expressed in this material are those of the author(s) and do not necessarily reflect the views of the National Science Foundation.

We thank Hemangi Murdande for her assistance during manual data labeling.
    
The authors have no competing interests to declare that are relevant to the content of this article.

\bibliographystyle{spmpsci}
\bibliography{bibliography}
\end{document}